\documentstyle[aps,prd,epsfig]{revtex}
\preprint{
}
\begin{document}
\draft

\title{Zero temperature string breaking in lattice quantum chromodynamics}

\author{Claude Bernard}
\address{
Department of Physics, Washington University, 
St.~Louis, MO 63130, USA
}
\author{Thomas DeGrand}
\address{
Department of Physics, University of Colorado, 
Boulder, CO 80309, USA
}
\author{Carleton DeTar and Pierre Lacock\thanks{
Present address: debis Systemhaus GEI, mbH, Pascalstra{\ss}e 8, 
52076 Aachen, Germnay}
}
\address{
Department of Physics, University of Utah Salt Lake City, 
     UT 84112, USA
}
\author{Steven Gottlieb}
\address{
Indiana University, Bloomington, IN 47405, USA
}
\author{Urs M.~Heller}
\address{
CSIT, The Florida State University, Tallahassee, FL 32306-4120, USA
}
\author{James Hetrick} 
\address{
University of the Pacific, Stockton, CA 95211, USA
}
\author{Kostas Orginos and Doug Toussaint}
\address{
Department of Physics, University of Arizona, Tucson, AZ 85721, USA
}
\author{Robert L.~Sugar}
\address{
Department of Physics, University of California, 
Santa Barbara, CA 93106, USA
}
\date{\today}
\maketitle
\begin{abstract}
The separation of a heavy quark and antiquark pair leads to the
formation of a tube of flux, or ``string'', which should break in the
presence of light quark-antiquark pairs.  This expected
zero-temperature phenomenon has proven elusive in simulations of
lattice QCD.  We study mixing between the string state and the
two-meson decay channel in QCD with two flavors of dynamical sea
quarks.  We confirm that mixing is weak and find that it decreases at
level crossing.  While our study does not show direct effects of
internal quark loops, our results, combined with unitarity, give clear
confirmation of string breaking.
\end{abstract}
\pacs{11.15.Ha, 12.38.Gc, 12.38.Aw}
%
%

\section{Introduction}
In the absence of dynamical sea quarks, the heavy quark-antiquark
potential is known quite accurately from numerical simulations of
lattice quantum chromodynamics \cite{ref:quenchedpot}.  The potential
is traditionally determined from the Wilson-loop observable, which is
proportional to $\exp[-V(R)T]$ at large $T$.  At large separation $R$,
the potential $V(R)$ rises linearly, as expected in a confining
theory.  In the presence of dynamical sea quarks the potential is
expected to level off at large $R$, signaling string breaking.  Thus
far, no SU(3) simulation at zero temperature with light sea quarks has
found clear evidence in the Wilson-loop observable for string breaking
\cite{ref:fullpot1,ref:fullpot2}, even out to $R \approx 2$ fm.

The reason string breaking has not been seen using the traditional
Wilson-loop observable is now clear
\cite{ref:michael,ref:DKKL,ref:drummond}.  The Wilson loop can be
regarded as a hadron correlator with a source and sink state (F)
consisting of a fixed heavy quark-antiquark pair and an associated
flux tube.  The correct lowest energy contribution to the Wilson-loop
correlator at large $R$ should be a state ``M'' consisting of two
isolated heavy-light mesons.  However, such a state with an extra
light dynamical quark pair has poor overlap with the flux-tube state,
so it is presumably revealed only after evolution to very large $T$.
To hasten the emergence of the true ground state, it is necessary to
enlarge the space of sources to include both F and at least one M
state.

Drummond demonstrated string breaking in a strong-coupling, hopping
parameter expansion with Wilson quarks
\cite{ref:drummond,ref:drummond2}.  A number of numerical studies of
theories less computationally demanding than QCD, including nonabelian
theories with scalar and adjoint matter fields, found string breaking
\cite{ref:Knechtli_Sommer,ref:Stephenson,ref:deForcrand_Philipsen,ref:Philipsen_Wittig,ref:Trottier}.
One study claims to have found string breaking in the absence of
dynamical sea quarks by doing a transfer matrix calculation
\cite{ref:Stewart_Koniuk}.

In a full SU(3) simulation, until recently, string breaking has only
been observed at nonzero temperature (close to, but below the
deconfinement crossover) \cite{ref:DKKL}, based on the Polyakov loop
observable, which evidently has much better overlap with the M state.
Some of us reported a preliminary low-statistics result for staggered
quarks in 1999 \cite{ref:lacock_lat99}, and, last year,
Pennanen and Michael announced evidence for string breaking at zero
temperature using Wilson-clover quarks and a novel technique for
variance reduction in computing the light quark propagator
\cite{ref:Pennanen_Michael}.  Duncan, Eichten, and Thacker found hints
of a flattening static potential using a truncated determinant
algorithm \cite{ref:DET}.

In this paper we demonstrate string breaking in an SU(3) simulation
with two flavors of dynamical sea quarks.  Our simulations are done in
the staggered fermion scheme on an archive of 198 configurations of
dimensions $20^3 \times 24$, generated with the conventional 
one-plaquette Wilson gauge action at $6/g^2 = 5.415$ and two flavors
of conventional dynamical quarks of mass $am = 0.0125$.  At this gauge
coupling and bare quark mass the lattice spacing is approximately
0.163 fm (based on a measurement of the Sommer
parameter\cite{ref:SOMMER} $r_0$ extracted from Wilson loops) with a
pi to rho mass ratio of 0.358.  These parameters were selected to give
a relatively light quark, making pair production energetically
favorable, and a large lattice volume (about 3.3 fm on a side and 3.9
fm in temporal extent) to allow ample room for string breaking.

Our computational methodology is described briefly in
Sec.~\ref{sec:method}.  In Sec.~\ref{sec:staggered} we justify our
fitting ansatz.  Finally, in Sec.~\ref{ref:results} we present our
results and conclusions.  Two Appendices describe our formalism for
random sources, and review the transfer matrix formalism we employ in
our analysis.

\section{Computational Methodology}
\label{sec:method}

Our conventional Wilson loop is computed with APE
smearing\cite{ref:APEblock} of the space-like gauge links.
Specifically, we used 10 iterations, combining the direct link with a
factor $1-\alpha$ (in our case, $\alpha = 0.294$) and six staples with
factor $\alpha/6$ with SU(3) projection after each iteration.  In
hamiltonian language the expectation value of this operator is the
correlator $G_{FF}(R,T)$ between an initial and final state $F$,
consisting of a static quark-antiquark pair separated by a fat string
of color flux.  Most of our results are obtained from on-axis Wilson
loops with $R$ ranging from 1 to 10, but we have two off-axis points
at displacement (2,2,0) and (4,4,0) (plus permutations and
reflections). Including other off-axis displacements might have been
statistically useful\cite{ref:fullpot2}.

We enlarge the source and sink space by including a meson-antimeson
state $M$ with an extra light quark located near the static antiquark
and an extra light antiquark, near the static quark.  To be precise,
this state is the tensor product of a static-light meson operator and
a static-light antimeson operator.  As discussed in Appendix B,
staggered flavor considerations make other choices more desirable
close to the continuum limit, e.g. the tensor product of creation
operators for a flux-tube state and a sigma meson.  Our static-light
meson construction makes the numerical analysis tractable and is
adequate for studying mixing on coarse lattices.  

We use an extended source for the light quark in the static-light
meson.  The heavy quark position, on the other hand, is fixed and used
to define the separation $R$.  Specifically, the gauge-invariant
source wavefunction at a site has support only on the site itself and
on the second on-axis neighbors in all six spatial directions,
connected to the central site by a product of the APE smeared links
along the paths.  Rather arbitrarily, the central site is given weight
2 while the satellite sites have weight 1.  Thus we also compute the
additional correlation matrix elements $G_{MM}(R,T)$, $G_{MF}(R,T)$
and $G_{FM}(R,T)$.  They are diagramed in Figs.~\ref{fig:D1} and
\ref{fig:E1}.

\figure{
 \epsfig{bbllx=50.9201,bblly=248.4926,bburx=565.2083,bbury=498.6348,clip=,
         file=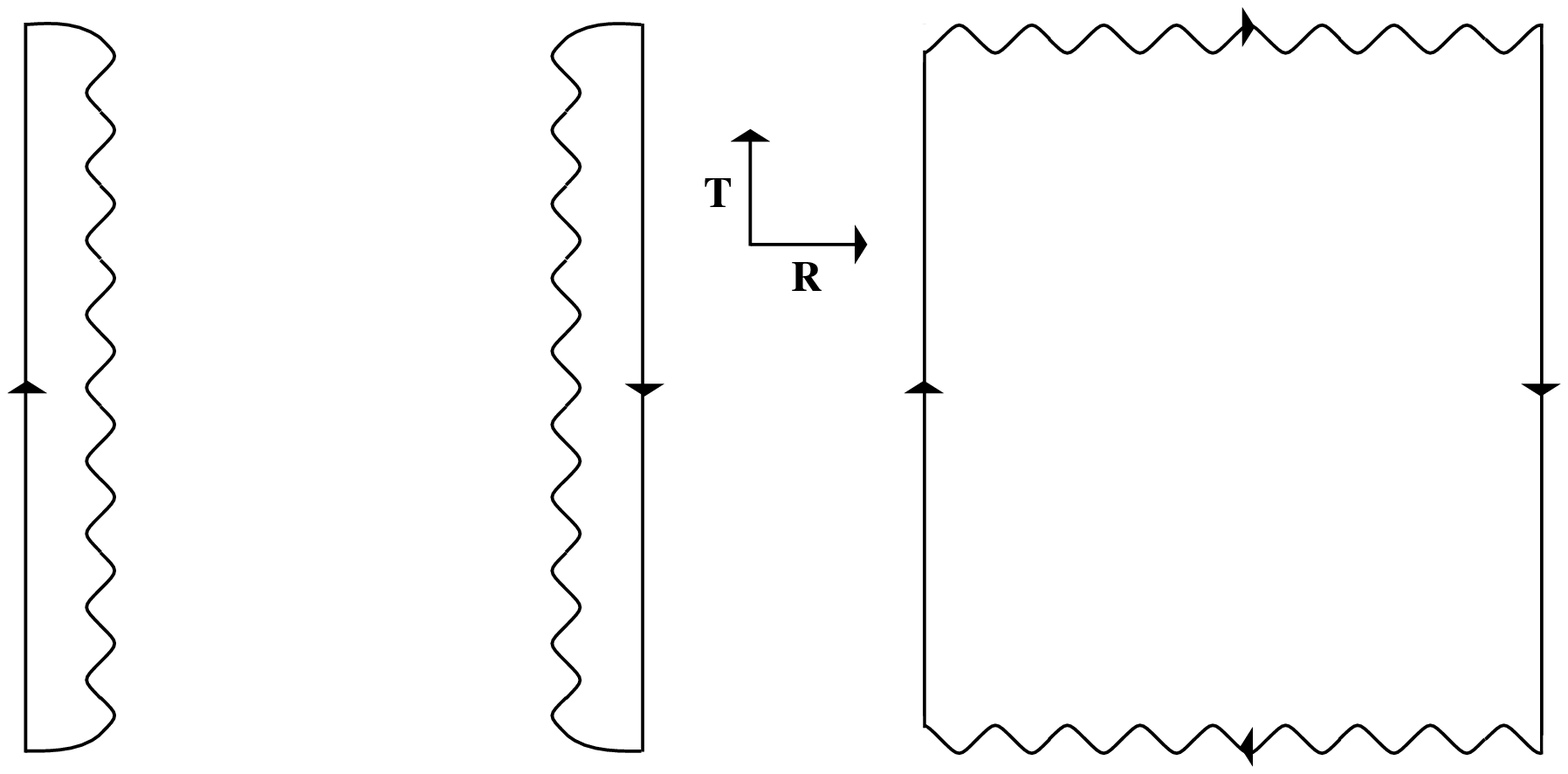,width=80mm}
\vspace*{2mm}
\caption{The static-light meson-antimeson pair contribution
to the full QCD propagator. The wiggly lines denote
the light quark propagator. Shown are the `direct' and 
`exchange' terms respectively.
\label{fig:D1}
}
}
\figure{
 \epsfig{bbllx=48.875,bblly=247.1593,bburx=565.2083,bbury=500.1324,clip=,
         file=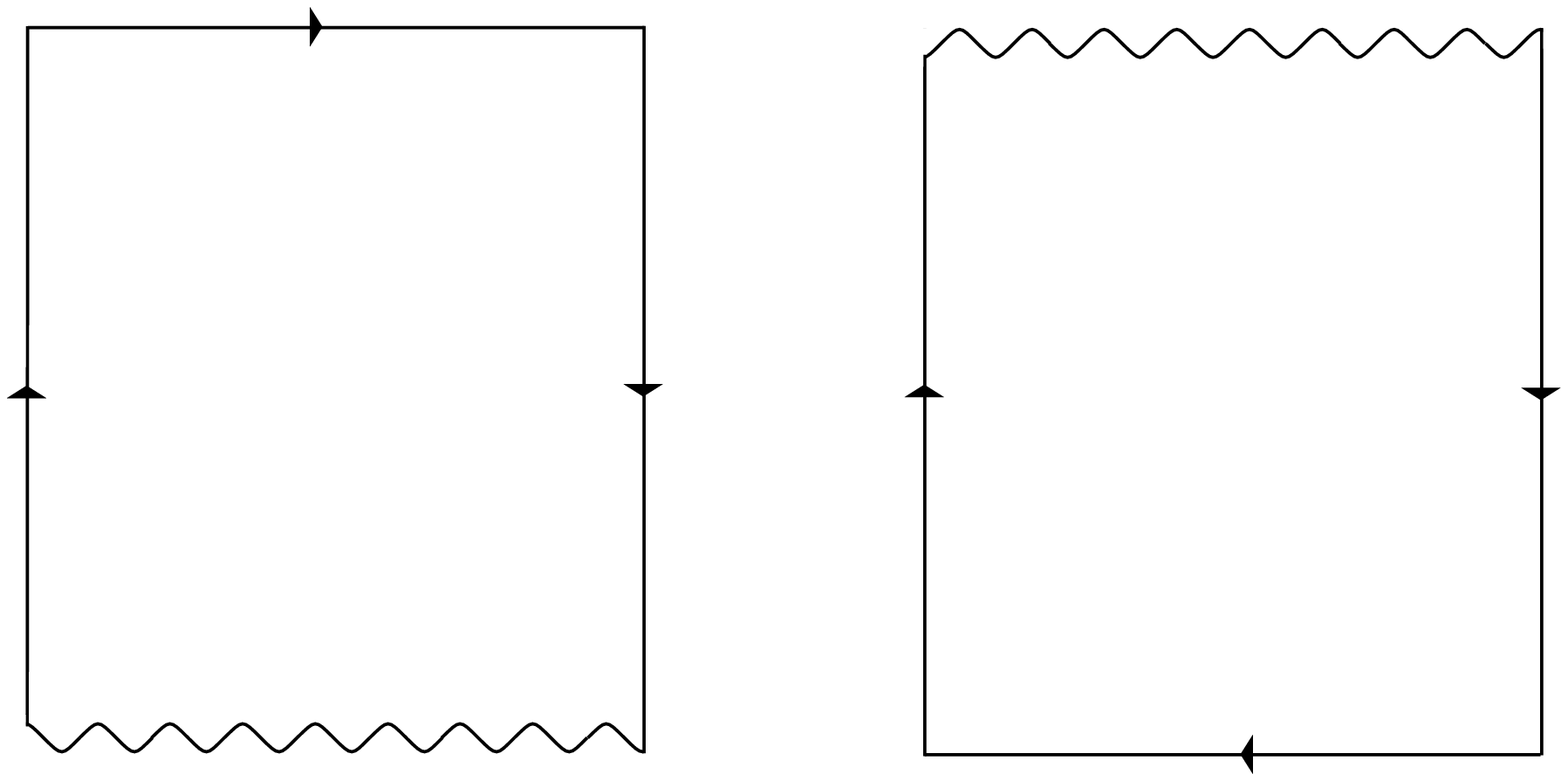,width=80mm}
\vspace*{2mm}
 \caption{The string-meson correlation matrix element $G_{FM}$
(and its hermitian conjugate $G_{MF}$).
The wiggly line again denotes the light quark propagator.
\label{fig:E1}
}
\vspace*{2mm}
}

For the light quarks in the static-light mesons we use the same
parameters as for the dynamical sea quarks.  To reduce variance we
generated ``all-to-all'' propagators for the light quark, using a
Gaussian random source method \cite{ref:Z2noise}. (See Appendix
\ref{sec:noise_appendix}).  Results reported here are based on 128
such sources per gauge configuration, which gave satisfactory
statistics for our lattice volume.  With this number of sources the
variance due to fluctuations in random source was comparable to that
due to fluctuations in the ensemble of gauge configurations.

We analyze our correlators using an extension of the transfer matrix
formalism of Sharatchandra, Thun, and Weisz\cite{ref:STW} for
staggered fermions, described in Appendix \ref{sec:tm_appendix}.  With
our choice of local meson operators, discrete lattice symmetries, 
also discussed in Appendix B,
require that all products of gauge links in the observables be
assigned phases consistent with being viewed as the paths of heavy
staggered fermions.  For example, for the Wilson-loop operator, a
hopping parameter expansion around an on-axis $R \times T$ rectangular
path gives, in addition to the conventional Wilson-loop gauge-link
product, a net phase factor $(-1)^{(R+1)(T+1)}$, independent of the
staggered fermion Dirac phase conventions.  (Included is a factor $-1$
for a single closed fermion loop.)  This phase then controls the sign
of the transfer matrix eigenvalue associated with the flux tube state.
We use a similar construction to get the phases for the nonclosed
gauge-link products in the diagrams of Figures \ref{fig:D1} and
\ref{fig:E1}.  In all cases the Dirac phase convention for the
gauge-link products must be consistent with that of the light quark.
A consequence of this construction is that the correlation
matrix elements involving off-axis gauge-link products must vanish
when summed over symmetry-equivalent paths for net displacements $\vec
R$ that have more than one odd Cartesian component.

\section{Staggered Fermion Pair Production}
\label{sec:staggered}

\subsection{Transfer matrix eigenvalues}

The heavy quark potential is defined as the ground state energy of the
QCD hamiltonian with a static quark and antiquark separated by
distance $R$.  Operationally, we extract the ground state energy by
fitting the time dependence of the correlation matrix elements in the
same manner as one obtains hadron masses.  Fundamentally, the
potential is determined by the eigenvalues of the transfer matrix.  To
justify our fitting ansatz, therefore, we start from an analysis of
the transfer matrix in the staggered fermion scheme.

Transforming to temporal axial gauge and making a suitable choice of
fermion phases, Sharatchandra, Thun, and Weisz showed that the
staggered fermion transfer matrix is hermitian but not positive
\cite{ref:STW}.  Then, it is convenient
to use the eigenvectors of the transfer matrix as a basis for
representing the correlation matrix.  In terms of the (possibly
negative) eigenvalues $\lambda_n(R)$ of the transfer matrix, our
correlation matrix can therefore be written in spectral
decomposition as
\begin{equation}
   G_{AB}(R,T) = \sum_{i=1}^N Z_{Ai}^*(R) Z_{Bi}(R)[\lambda_i(R)]^{T+1} ~,
\label{eq:transition}
\end{equation}
where $A,B$ refer to the flux tube $F$ or meson-meson $M$ states.
The $T+1$ power is natural, as we show in Appendix \ref{sec:tm_appendix}.
This result forms the basis for our fitting ansatz.  To apply this
decomposition to our results, it is essential, as we have done, that
we treat the heavy quark lines as static staggered quarks, with all
fermion phases included, and that source and sink operators are
equivalent.

\subsection{Static light propagator}

We first examine the single static-light meson correlator, shown in
Fig.~\ref{fig:slprop}.  We find good fits to two spectral components,
one with no phase oscillation in $T$, corresponding to a $S$-wave
light quark and a positive transfer-matrix eigenvalue, and the other,
higher in energy, and with oscillating phase in $T$, corresponding to
a $P$-wave light quark and a negative transfer-matrix eigenvalue.
Fitting to a single nonoscillating exponential plus a single
oscillating exponential over the range $t = [2,9]$ gives energies
(defined as usual as $\log(|\lambda|)$) $aE_S = 0.7884(12)$ and $aE_P
= 1.022(6)$ with $\chi^2/df = 2.7/4$.  The $P$-wave amplitude is
suppressed by a factor of about 0.5 relative to the $S$-wave
amplitude.

\figure{
 \vspace*{-2.5cm}
 \epsfig{bbllx=200,bblly=130,bburx=830,bbury=940,clip=,
         file=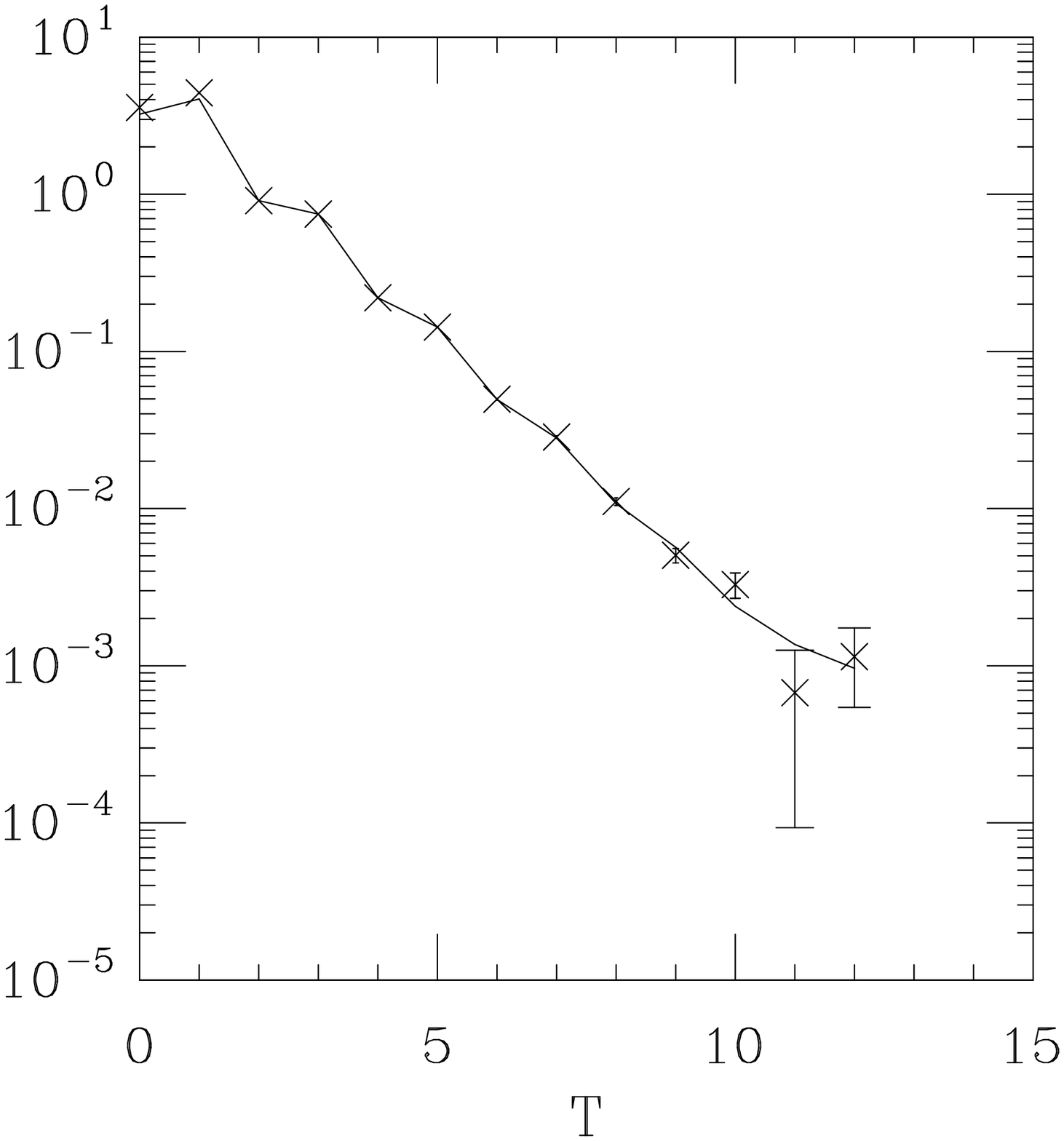,width=70mm}
\caption{Static light propagator with a nonoscillating $S$-wave and
oscillating $P$-wave component. The solid line connects the best fit
values.  \label{fig:slprop}
}
}
\subsection{Fitting form for the correlation matrix}

Since we expect to find two static-light mesons at large $R$ in our
$G_{MM}$ correlator, we look for a positive eigenvalue ``SS'' spectral
component corresponding to two $S$-wave mesons and a negative
eigenvalue SP component corresponding to an $S$-wave and $P$-wave
meson.  For most of our analysis we omit the PP component, a choice
that we justify as follows: Squaring the static light propagator
suggests that the PP channel would contribute a second nonoscillating
spectral component at a higher energy (about 0.5 in lattice units)
than SS and with a smaller amplitude (about 0.2).  We present results,
however, that include a excited state component inspired by the PP
contribution.  The SS and SP components would be expected to have a
smooth dependence on $R$ for large $R$.  To these two spectral
components we add a third, corresponding to a conventional Wilson-loop
contribution at short distance $R$.  With the staggered fermion phases
included, the net Wilson-loop phase factor $(-1)^{(R+1)(T+1)}$
produces a transfer matrix eigenvalue with a phase $(-)^{R+1}$ that
oscillates with $R$.


Our proposed fitting ansatz is thus Eq (\ref{eq:transition}) with $N =
3$ and with the explicit SS, SP, and flux-tube eigenvalues
(respectively)
%
%
\begin{eqnarray}
\lambda_1(R) &=& e^{-V_1(R)} \nonumber \\
\lambda_2(R) &=& -e^{-V_2(R)}  \label{eq:fit_ansatz} \\
\lambda_3(R) &=& (-)^{R+1}e^{-V_3(R)} ~. \nonumber 
\end{eqnarray}
Other components can be readily included.  With our choice of sources
and sinks the correlation matrix is found to be real, so we may take
real $Z$ factors.  An ambiguity permits changing the sign
simultaneously in $Z_{Fi}$ and $Z_{Mi}$, which we resolve arbitrarily
by requiring $Z_{Mi}$ to be positive.  At large $R$ we expect $V_1(R)$
to approach $2aE_S$ and $V_2(R)$ to approach $aE_S + aE_P$ and at
small $R$ we expect $V_3(R)$ to correspond roughly to a Coulomb plus
linear heavy quark potential.  At intermediate $R$ we expect mixing
among these states.  Avoided level crossing may occur at even $R$
between intermediate states 2 and 3 and at odd $R$ between
intermediate states 1 and 3.

\section{Results}
\label{ref:results}

\subsection{Potential}

With the factorization inherent in our ansatz (\ref{eq:transition})
and the choice $N=3$, we are fitting three correlators with nine
parameters.  Our fitting range in $T$ varies over the data set as
shown in Table \ref{tab:resultsa} with typically 10 or more degrees of
freedom.  The goodness of fit supports our ansatz.

Our selection of fit ranges compromised between the need to get
acceptable fits and our intention to vary the end points $t_{\rm min}$
and $t_{\rm max}$ smoothly as a function of $R$.  At low $R$ the
$G_{FF}$ (flux-tube-type) correlator has quite small errors, whereas
at larger $R$, errors increase, especially at higher $T$.  Thus for
small $R$ we can set a higher $t_{\rm min}$ in the $G_{FF}$ correlator
to improve chi square without loss of information.  At larger $R$
there seems to be little improvement in goodness of fit in setting a
high $t_{\rm min}$, and the large statistical errors at high $T$ give
no advantage to setting a higher $t_{\rm max}$.

To give an impression of the quality of the fits, we plot the {\em
absolute value} of the correlation matrix elements vs $T$ for two
values of $R$ in Figs.~\ref{fig:r3corr} and \ref{fig:r6corr}.  Also
plotted are the absolute values of the fitting functions.  The errors
on the observed $G_{FM}$ and $G_{MM}$ correlators also give an
indication of the signal obtainable with the random source method.

\figure{
 \vspace*{-3cm}
 \epsfig{bbllx=200,bblly=130,bburx=830,bbury=940,clip=,
         file=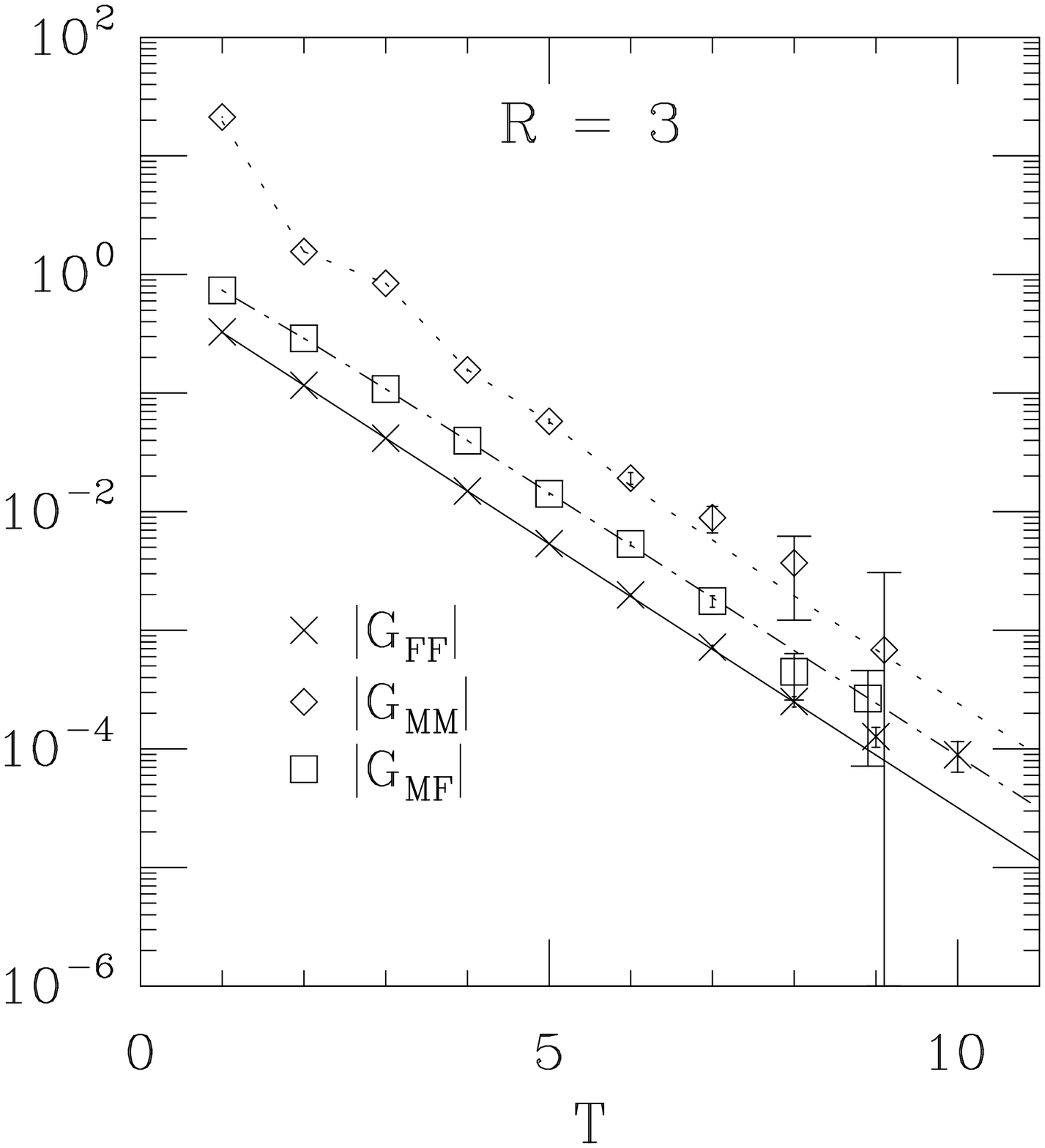,width=80mm}
\caption{Absolute value of the correlation matrix elements vs $t$ at
$R=3$.  The lines connect the best fit values.  The fit ranges are
$[4,9]$ for $G_{FF}$, $[1,9]$ for $G_{FM}$ and $[2,9]$ for $G_{MM}$.
\label{fig:r3corr}
}
}

\figure{
 \vspace*{-3cm}
 \epsfig{bbllx=200,bblly=130,bburx=830,bbury=940,clip=,
         file=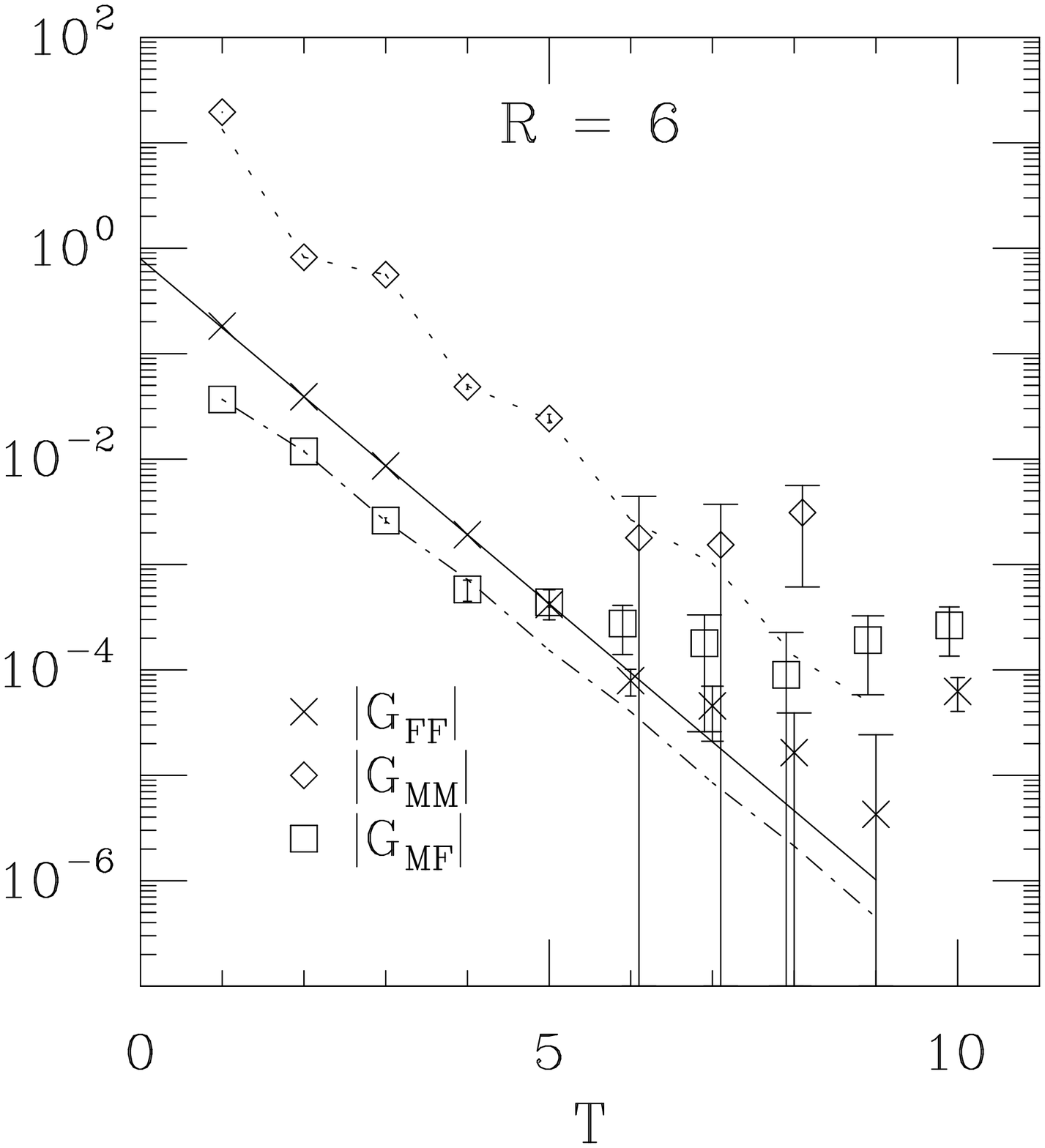,width=80mm}
 \caption{Same as Fig.~\protect\ref{fig:r3corr}, but with $R=6$. 
The fit ranges are $[2,9]$ for $G_{FF}$, $[1,7]$ for 
$G_{FM}$, and $[2,7]$ for $G_{MM}$.
\label{fig:r6corr}
}
}

Results at $R=3$ are a good representative of the small $R$
correlators and the degree to which the fit results are affected by
the choice of end points.  Fitting the three correlators $G_{FF}$,
$G_{FM}$, and $G_{MM}$ over the ranges $T \in [6,9]$, $[2,9]$, and
$[1,9]$, respectively, gave $V_1 = 1.60(5)$, $V_2 = 1.89(11)$, and
$V_3 = 1.026(6)$ with $\chi^2/df = 14.6/12$. Changing the fit ranges
to $t \in [2,9]$, $[2,9]$, and $[2,9]$ increased $\chi^2/df$ to
23.6/15 and gave $V_1 = 1.62(5)$, $V_2 = 1.88(11)$, and $V_3 =
1.025(8)$.

To see the effect upon the mixing analysis of including other states,
we experimented with adding an excited state modeled after the
two-meson PP spectral component, which contributes in the same way as
the SS component.  The fourth spectral component is denoted $V_4(R)$.
Doing so increases the parameter count to 12.  To assure stability of
the fits, we fixed the two-meson energies $V_1(R) = 2E_S$ and $V_4(R)
= 2E_P$, leaving 10 free parameters.  We found acceptable fits.  The
unconstrained energies $V_2$ and $V_3$ agreed within errors with
results from the three-spectral-component ansatz.  For example at $R =
3$ over fit ranges $t \in [4,9]$, $[1,9]$, $[1,9]$ we find $\chi^2/df
= 22.9/14$ with $V_2 = 1.95(2)$ and $V_3 = 1.024(4)$.

Our fit results for the three-spectral-component ansatz are listed
in Tables \ref{tab:resultsa} and \ref{tab:resultsb} and plotted in
Fig.~\ref{fig:potential}.  For small distances the flux tube energy
$V_3(R)$ is smallest and the flux tube state dominates the large $T$
behavior of the correlation matrix, while at large distances the
two-meson energies $V_1(R)$ and $V_2(R)$ are smaller and the two-meson
states dominate the correlation matrix at large $T$.  With our choice
of light quark mass the first level crossing occurs at $R = 6a$, or
0.98 fm.  The string is broken.  It is interesting that the energies
$V_1(R)$ and $V_2(R)$ are very nearly equal to their asymptotic values
throughout.  Thus we see no spectral evidence of a meson-meson
interaction at the level of our statistics.

It is clear from these results that mixing between the flux-tube and
two-meson channels is weak.  There is no evident rounding of the
potentials normally associated with avoided level crossing.  At higher
order in mixing we would expect to require two-meson spectral
components in the flux tube $G_{FF}$ correlator.  They should appear
as a result of the breaking and rejoining of the string.  However, the
amplitudes for both terms in this correlator are small enough that, if
it weren't for the enforcement of a common spectrum and factorization
in our fit ansatz, they might have been missed.  The converse presence
of the string term in the diagonal two-meson correlator $G_{MM}$ can
be accounted for by the ``box'' diagram in the quark correlator that
resembles a Wilson loop.

As a check of mixing between the flux-tube level and the two-meson
levels, we examined the transition amplitude $G_{FM}$ to see if, by
itself, it contained both types of spectral components
\cite{ref:Aoki}.  To do so we carried out a separate three-exponential
fit to the transition amplitude $G_{FM}$ alone, fixing the energies of
the $SS$, $SP$, and flux-tube spectral components to the values found
in the multichannel channel analysis, but adjusting their amplitudes
for a best fit.  For $r \le 5$, where the spectral components are
clearly nondegenerate, we found that the amplitudes for the $SS$ and
flux-tube components were both nonzero at the three to five-sigma
level.  Thus our results, combined with unitarity, confirm mixing and
imply string breaking.

We have required that the valence and sea quarks match, thus assuring
that the correlation matrix is a power of the transfer matrix.
However, if mixing effects are very weak, had we chosen, instead, to
omit the dynamical sea quarks altogether, it would very likely be
difficult to detect the consequent inconsistencies.  We analyze mixing
further in the following subsection.

\figure{
 \vspace*{-4cm}
 \epsfig{bbllx=200,bblly=130,bburx=830,bbury=940,clip=,
         file=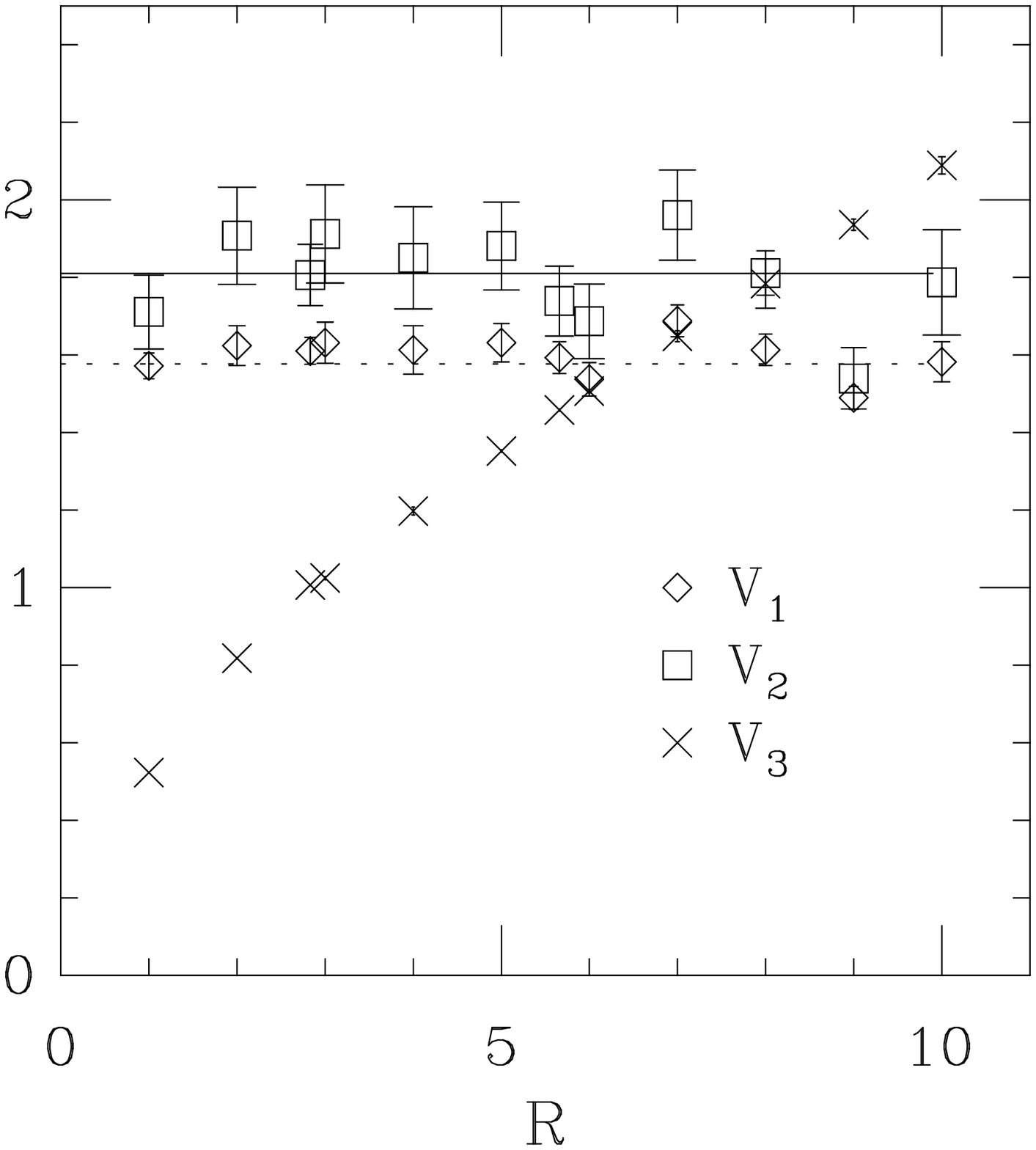,width=120mm}
 \caption{Heavy quark potential and first two excited states {\it vs}
separation $R$.  The dashed and solid lines give the asymptotic values
$2aE_S$ and $E_P + E_S$.  Jackknife errors are shown.
\label{fig:potential}
}
}

\subsection{Modeling the mixing}
\label{sec:mixingmodel}

Drummond, Pennanen and Michael 
analyze their results in terms of a transfer matrix model 
that mixes the two-meson and flux-tube states
\cite{ref:drummond,ref:Pennanen_Michael}.  Our approach differs
slightly, because our multi-exponential fit carries more information
and because our multi-channel contributions complicate the analysis.
A perturbative model of string breaking starts with a zeroth order
pure flux-tube state ($F^0$) and a pure two-meson state, consisting of
either a pair of unperturbed $S$-wave mesons ($S^0S^0$) or unperturbed
$S$- and $P$-wave mesons ($S^0P^0$).  (By extension, we could include
the $P^0P^0$ channel.)  At a given $R$ the lattice mixing between the
states can be described by a transfer matrix on the unperturbed basis
\begin{equation}
    {\cal T}(R) = \left(\begin{array}{ccc}
        \lambda^0_1(R) & 0 & x \\
        0 & \lambda^0_2(R) & y \\
        x & y & \lambda^0_3(R)  
    \end{array} \right)
\end{equation}
that evolves a state across a Euclidean time slice.  The rows and
columns are arranged in the order $S^0S^0$, $S^0P^0$, and $F^0$. We
assume a small value for the mixing $x$ between $F^0$ and $S^0S^0$ and
$y$ between $F^0$ and $S^0P^0$.  Although the $S^0S^0$ and $S^0P^0$
states may mix, for simplicity, we have ignored this effect.  The
diagonal elements correspond to our conventions for our fit ansatz
(\ref{eq:fit_ansatz}).  The $2 \times 2$ correlation matrix connecting
our flux-tube and two-meson source and sink states at a given $R$ is
\begin{equation}
   G(R,T) = {\tilde Z}^0(R) {\cal T}(R)^{T+1} Z^0(R) ~,
\end{equation}
where $Z^0$ is the unperturbed $2 \times 3$ matrix used in our fit
ansatz.  The diagonal elements of $G(T)$ and the potentials are
unperturbed at first order in $x$ and $y$.  The off-diagonal
correlator to first order is
\begin{equation}
  G_{FM}(R,T) = A_1 [\lambda^0_1(R)]^{T+1} +  A_2 [\lambda^0_2(R)]^{T+1}
                         +  A_3 [\lambda^0_3(R)]^{T+1} ~,
\end{equation}
where
\begin{eqnarray}
    A_1 &=& Z_{F1}(R) Z_{M1}(R) =
        Z^0_{F3}(R) Z^0_{M1}(R) \frac{x}{\lambda^0_1(R) - \lambda^0_3(R)} 
      \nonumber \\
    A_2 &=& Z_{F2}(R) Z_{M2}(R) =
        Z^0_{F3}(R) Z^0_{M2}(R) \frac{y}{\lambda^0_2(R) - \lambda^0_3(R)} 
       \\
    A_3 &=& Z_{F3}(R) Z_{M3}(R) = - (A_1 + A_2) \nonumber 
\end{eqnarray}
To first order in the mixing parameters, the $Z$ factors are
unperturbed, and we may equate $Z = Z^0$ to obtain three constraints
for the two mixing parameters $x$ and $y$.  The third constraint is a
sum rule.  To apply the model we chose to impose the sum rule as an
{\it a posteriori} constraint on the parameters of the fit for each
$R$:
\begin{equation}
   f(Z) =  Z_{F1}Z_{M1} + Z_{F2}Z_{M2} + Z_{F3}Z_{M3} = 0 ~,
\end{equation}
and use the first two conditions to determine the mixing parameters:
\begin{eqnarray}
   x &=& [\lambda^0_1(R) - \lambda^0_3(R)] Z_{F1}/Z_{F3} \nonumber  \\
   y &=& [\lambda^0_2(R) - \lambda^0_3(R)] Z_{F2}/Z_{F3} ~.
\label{eq:mixing}
\end{eqnarray}
This was done by linearizing the sum rule in the vicinity of the
minimum $Z_*$ of the unconstrained $\chi^2$:
\begin{equation}
   f(Z_0) = 0 = f(Z_*) + \nabla f(Z_*) \cdot (Z_0 - Z_*) ~.
\end{equation}
It is straightforward to determine the attendant increase in $\chi^2$,
the shift in parameters, and the decrease in errors.  Since this
procedure assumes the sum rule can be linearized, it is valid only to
the extent that the increase in $\chi^2$ is small.

Table \ref{tab:mixing_three} lists results.  Shown are the values of the
coefficient ratio 
\begin{equation}
   d = -(Z_{F1}Z_{M1} + Z_{F2}Z_{M2})/Z_{F3}Z_{M3} ~,
\end{equation}
before and after imposing the linearized sum rule constraint as well
as the shifts in $\chi^2$ and the values of $x$ and $y$ obtained after
imposing the constraint.  The ratio $d$ should be one if the
sum rule is satisfied.  We see that the increase in $\chi^2$ is
smallest for $R \ge 5.66$, but it is otherwise unacceptably large.
Thus the mixing model suits our three-exponential ansatz only for
larger $R$.  The agreement improves considerably when we include the
two-meson PP spectral component, as discussed above, and fix the
two-meson energies $V_1(R) = 2E_S$ and $V_4(R) = 2E_P$, leaving 10
free parameters.  Although we include the fourth component in the fit,
we still consider only a three-state mixing model.  (In effect, we
have set mixing to the fourth level to zero.)  Results are shown in
Table \ref{tab:mixing_four}.  Now the mixing model seems plausible for
$R \ge 2$.  

The mixing model makes separate predictions for the connected and
disconnected meson-to-meson correlators, which provides an additional
constraint on the mixing parameters.  As a test of the systematic
error arising from model assumptions, we have tried refitting all of
our data to a purely four-component mixing model, with separate
disconnected and connected correlators.  While the resulting mixing
coefficients repeat the trends of Tables \ref{tab:mixing_three} and
\ref{tab:mixing_four}, the values differ as much from those of the
tables as the two tables do from each other.

\figure{
 \vspace*{-5cm} \hspace*{-1cm}
 \epsfig{bbllx=200,bblly=130,bburx=1030,bbury=940,clip=,
         file=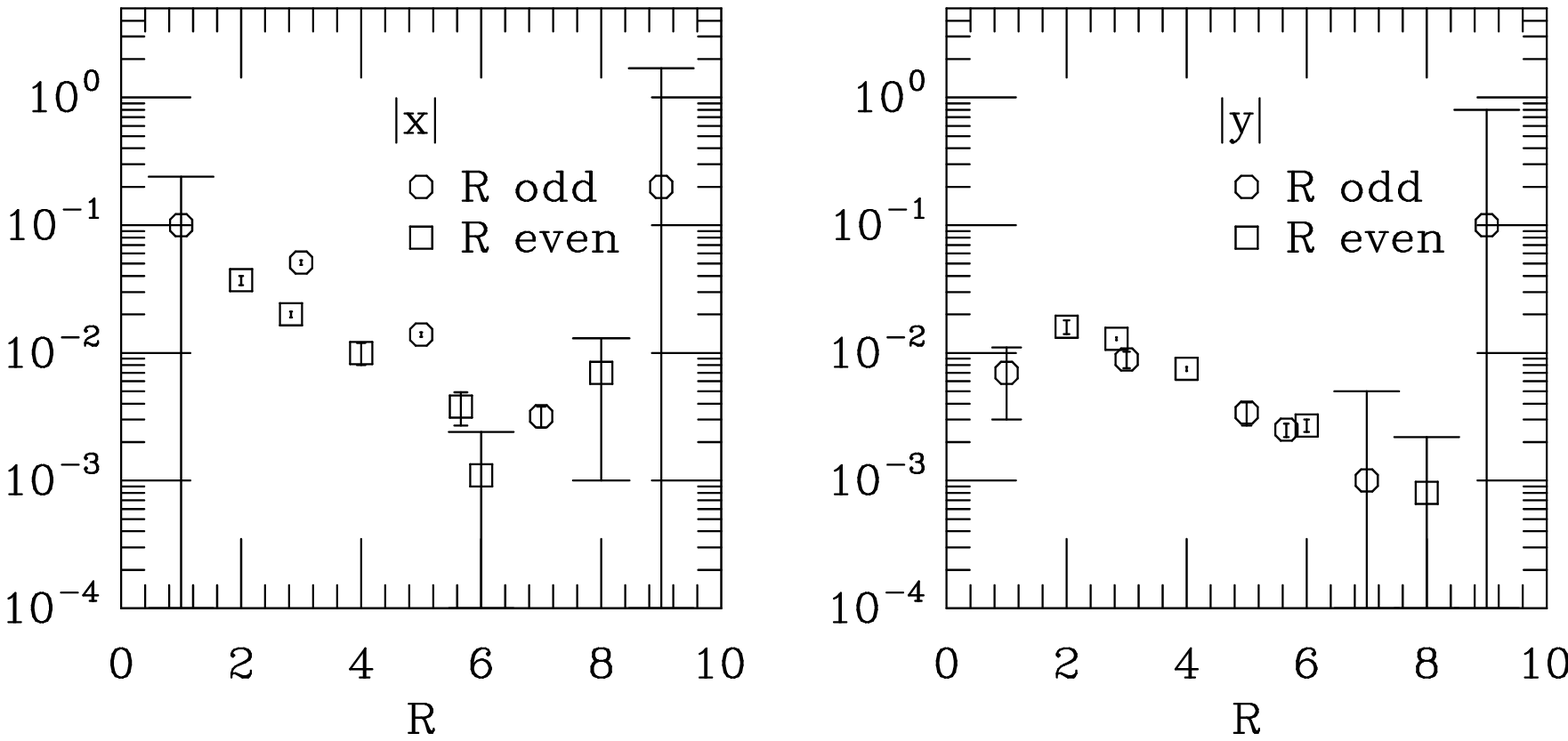,width=160mm}
 \vspace*{-1cm}
\caption{Absolute value of the mixing parameters $x$ and $y$ 
{\it vs} separation $R$.  Odd and even series are distinguished.
\label{fig:mixing} 
}
\vspace*{5mm}
%

If we now take the mixing model at face value, it is interesting to
consider how the strength of mixing varies with $R$.  In
Fig.~\ref{fig:mixing} we see a significant decrease in both $x$ and
$y$ with increasing $R$, which is to be expected partly from
(\ref{eq:mixing}) as the eigenvalues cross.  There is also a
pronounced difference in mixing strengths at even and odd $R$,
suggesting a suppression of mixing between unperturbed oscillating and
nonoscillating levels.  Obviously, given the coarseness of our
lattice, the comparison with a model Hamiltonian must be done
judiciously.  However, even these first, crude QCD results should help
constrain the phenomenological analysis of quarkonium decay
\cite{ref:Drummond_Horgan}.

\section{Conclusions} 

We have studied string breaking in the heavy quark interaction at zero
temperature in QCD.  Our calculation used two flavors of light
staggered quarks and gauge configurations generated in the presence of
the same quarks.  We extended the analysis of the spectrum of the
transfer matrix in the staggered fermion formalism to treat our
nonlocal sources.  By adding explicit two-meson states to the
conventional flux-tube state, we obtain the expected result that the
two-meson state is energetically favored at large distance.  Our
two-channel correlators fit a model with three factorizing spectral
components and a partially constrained, extended model with four
spectral components.  Our results are also consistent with a simple
three- and four-state model with weak mixing.  The mixing coefficients
in the mixing model matrix appear to decrease at the level crossing
points.  Our principal finding is that mixing between the flux-tube
and two-meson channels is indeed weak.  Thus we see why string
breaking has been missed in the flux-tube channel by itself.  While
our matching of dynamical and valence quarks is designed to satisfy
unitarity, with our statistics we have not found compelling evidence
for quark loop effects.  Doubtless, our results could have just as
well been reproduced in a quenched simulation, where
inconsistencies with unitarity would appear only at higher order in
the mixing matrix elements.  However, we also find that the transition
correlator by itself connecting flux-tube and two-meson channels is
nonzero for all $r$ and shows both string-like and two-meson-like
spectral components, at least for $r < 6$. Thus, our results, combined
with unitarity, require string breaking.  It is recommended that
future staggered fermion studies closer to the continuum take care to
restrict the two-meson wave function to the light-flavor singlet
channel.

\acknowledgments 

Gauge configurations were generated on the Indiana University Paragon.
Calculations were carried out on the IBM SP at SDSC and the IBM SP and
Linux cluster at the University of Utah Center for High Performance
Computations. This work was supported by the U.S. Department of Energy
under grants
DE$-$FG02$-$91ER$-$40661, 
DE$-$FG02$-$91ER$-$40628, 
DE$-$FG03$-$95ER$-$40894, 
DE$-$FG03$-$95ER$-$40906, 
DE$-$FG05$-$96ER$-$40979,  
and the National Science Foundation under grants
PHY99$-$70701 and     
PHY97$-$22022.        

\appendix
\section{Random Source Estimators Applied to String Breaking}
\label{sec:noise_appendix}

We use random source methods to calculate the all-to-all quark
propagators in this study.  Here we outline the method as applied to
string breaking.  We begin with the conventional light quark
propagator
\begin{equation}
G_q^{ab}(x,y) = \langle q^a(x) \bar q^b(y) \rangle ~,
\end{equation}
which satisfies the equation
\begin{equation}
( {D}\!\!\!\!/\, + m {\mathbf 1} )^{ab}_{x,y} G_q^{bc}(y,z) =
 \delta^{ac} \delta_{x,z} ~.
\end{equation}
Summation over repeated indices is implied unless noted otherwise.
${D}\!\!\!\!/\,$ denotes the usual staggered lattice Dirac operator.

The static (infinitely heavy) quark propagator can be found from an
expansion to leading order in $1/m$.  The result depends on whether
propagation is forward or backward in time.  For $t > t^\prime$,
omitting the heavy quark mass factors,
\begin{equation}
G_h^{ab}[(\vec x,t),(\vec y,t^\prime)] =
 \langle h^a(\vec x,t) \bar h^b(\vec y,t^\prime) \rangle =
 \delta_{\vec x, \vec y} \left( \prod_{\tau=t-1}^{t^\prime}
 \tilde U^\dagger_4(\vec x,\tau) \right)^{ab} ~,
\end{equation}
where
\begin{equation}
\tilde U^\dagger_\mu(\vec x,\tau) = \alpha(\vec x,\tau;\mu)
     U^\dagger_\mu(\vec x,\tau)
\label{eq:link_with_phases}
\end{equation}
and the $\alpha(\vec x,\tau;\mu)$ are staggered fermion phase factors.
For $t < t^\prime$
\begin{equation}
G_h^{ab}[(\vec x,t),(\vec y,t^\prime)] = 
  \delta_{\vec x, \vec y} \left( \prod_{\tau=t}^{t^\prime-1}
  \tilde U_4(\vec x,\tau) \right)^{ab} ~,
\end{equation}
where
\begin{equation}
  \tilde U_\mu(\vec x,\tau) = -\alpha(\vec x,\tau;\mu)U_\mu(\vec x,\tau) ~.
\end{equation}
We will be interested in static-light mesons, which we construct from the 
Grassman fields as
\begin{eqnarray}
B(\vec x,t) &=& \sum_{\vec y} \sum_{a,b}
 \bar h^a(\vec x,t) \rho_t^{ab}(\vec x, \vec y)
 q^b(\vec y,t) ~, \nonumber \\
\bar B(\vec x,t) &=&  \sum_{\vec y} \sum_{a,b}
 \bar q^a(\vec y,t) \rho_t^{ab}(\vec y, \vec x)
 h^b(\vec x,t) ~.
\end{eqnarray}
The wave function $\rho_t$ is hermitian, $\rho_t^\dagger = \rho_t$, and
depends via gauge fields on $t$.  We use
\begin{equation}
\rho_t^{ab}(\vec x, \vec y) = 2 \delta^{ab} \delta_{\vec x, \vec y} +
 \sum_{\mu=1}^3 \left\{ \left( \prod_{l=0}^{r_0-1}
 \tilde U_\mu(\vec x + l \hat \mu, t) \right)^{ab}
 \delta_{\vec x, \vec y - r_0 \hat \mu} + \left( \prod_{l=1}^{r_0}
 \tilde U_\mu^\dagger(\vec x -l \hat \mu, t) \right)^{ab}
 \delta_{\vec x, \vec y + r_0 \hat \mu} \right\} ~.
\end{equation}
The space-like gauge-link matrices can be taken as (APE) smeared gauge
fields.  Thus the tilde here represents both smearing and the
inclusion of the fermion hopping phases.  On the other hand the
time-like gauge-link matrices are not smeared.

For the static-light meson correlation function with source at the
origin we obtain
\begin{eqnarray}
G_B(T) &=& \langle B(\vec 0, T) \bar B(\vec 0, 0) \rangle = 
 G_h^{ab}[(\vec 0, 0), (\vec 0, T)] \rho_T^{bc}(\vec 0, \vec x)
 G_q^{cd}[(\vec x,T),(\vec y,0)] \rho_0^{da}(\vec y, \vec 0) \nonumber \\
 &=& \langle \eta^{\dagger a}(\vec 0,0)
 V^{ab}[(\vec 0,0),(\vec 0,T)]
 \rho_T^{bc}(\vec 0, \vec x) G_q^{cd}[(\vec x,T),(\vec y, t^\prime)] 
 \rho_{t^\prime}^{de}(\vec y, \vec z) \eta^e(\vec z, t^\prime)
 \rangle_\eta ~.
\label{eq:SLC}
\end{eqnarray}
where
\begin{equation}
 V^{ab}[(\vec x,0),(\vec x,T)] = \left( \prod_{\tau=0}^{t-1}
 \tilde U_4(\vec x, \tau)] \right)^{ab}
\end{equation}
is the product of link matrices and hopping phase factors from $(\vec
x,0)$ to $(\vec x,T)$.  In the last line we introduced Gaussian random
numbers, $\langle \eta^{\dagger a}(x) \eta^b(y) \rangle_\eta =
\delta^{ab} \delta_{x,y}$, to compute the trace. Multiplied by
$\rho_t$ they form the source for the light quark inversion and are
then used again in the computation of the static-light meson
correlation function as seen in eq.~(\ref{eq:SLC}).  Introducing the
smeared, random source propagator
\begin{equation}
\Psi_\eta^a(\vec x,t) = \rho_t^{ab}(\vec x, \vec y)
 G_q^{bc}[(\vec y,t),(\vec z, t^\prime)] \rho_{t^\prime}^{cd}(\vec z, \vec u)
 \eta^d(\vec u, t^\prime) ~,
\label{eq:fuz_prop}
\end{equation}
we can write the static-light meson correlation function as
\begin{equation}
G_B(T) = \langle \eta^{\dagger a}(\vec 0,0) V^{ab}[(\vec 0,0),(\vec 0,T)]
 \Psi_\eta^b(\vec 0,T) \rangle_\eta ~.
\end{equation}
In numerical practice this result is computed for a source at any
lattice location and averaged over the space-time volume.

For the computation of the heavy quark potential and the investigation
of string breaking we will need a heavy quark-antiquark ``string state''
\begin{equation}
O_F(\vec x, \vec y, t) = \bar h^a(\vec x, t) V^{ab}[(\vec x,t),(\vec y, t)]
h^b(\vec y, t)/\sqrt{N_c} ~,
\end{equation}
where $V^{ab}[(\vec x,t),(\vec y, t)]$ is a superposition of products of (APE
smeared) gauge fields in time slice $t$ starting at $\vec x$ and ending
at $\vec y$.

We are then interested in the diagonal correlator between a string state
at time $0$ and one at time $T$ $-$we consider the connected part only
\begin{equation}
\langle O_F(\vec R, \vec 0, T) O_F(\vec 0, \vec R, 0) \rangle = W(\vec R, T) ~,
\end{equation}
with smeared space-like gauge field product segments in the time-like
Wilson loop $W(\vec R, T)$.  Except for the fermion hopping phases,
which give a net factor $(-)^{(R+1)(T+1)}$ independent of staggered
fermion phase convention, this is the correlation function usually
considered for the computation of the heavy quark potential.  As
usual, in practice this quantity is averaged over all choices of
lattice origins and on-axis displacements $R$.  We also computed it
for two off-axis displacements, namely $R = \sqrt{8}$ and $2\sqrt{8}$.
For off-axis displacements $V^{ab}[(\vec x,t),(\vec y, t)]$ is
constructed from a symmetric set of space-like paths joining the
endpoints.

We are also interested in the off-diagonal correlator between the
string state and the two-meson state
\begin{equation}
   O_M(\vec 0, \vec R, t) = \bar B(\vec R, t) B(\vec 0, t) ~.
\end{equation}
\begin{eqnarray}
\label{eq:str_BB}
&& \langle O_F(\vec R, \vec 0, T) O_M(\vec 0, \vec R, 0) \rangle
 \nonumber \\
 &=& V^{ab}[(\vec x,0),(\vec x,T)]
 V^{bc}[(\vec R,T),(\vec 0, T)]
 V^{cd}[(\vec R,T),(\vec R,0)]
 \rho_0^{de}(\vec 0, \vec x)
 G_q^{ef}[(\vec x,0),(\vec y,0)] \rho_0^{fa}(\vec y, \vec R) \nonumber \\
 &=& \langle \eta^{\dagger a}(\vec R,0)
 V^{ab}[(\vec R,0),(\vec R,T),(\vec 0,T),(\vec 0,0)]
 \Psi_\eta^b(\vec 0,0) \rangle_\eta ~.
\end{eqnarray}
The $V$ with four coordinate arguments stands for the product of
time-like and (APE smeared) space-like gauge fields that make up the
part of a time-like Wilson loop without the spatial segment at time
$t=0$.  Again, this correlator is computed with the Gaussian random
source method, in a manner similar to the static-light meson
correlation function~(\ref{eq:SLC}).  The other off-diagonal
correlator, $\langle B(\vec R, T) \bar B(\vec 0, T) S(\vec 0, \vec R,
0) \rangle$, is the hermitian conjugate of (\ref{eq:str_BB}).  This
quantity is averaged over the same choices of origin and displacement
as the Wilson loop.

Finally, we need the two-meson correlator
\begin{eqnarray}
&& \langle O^\dagger_M(\vec 0, \vec R, T) O_M(\vec 0, \vec R, 0)
 \rangle \nonumber \\
 &=&  V^{ab}[(\vec R,0),(\vec R,T)]
 \rho_T^{bc}(\vec R, \vec x) G_q^{cd}[(\vec x,T),(\vec y,T)]
 \rho_T^{de}(\vec y, \vec 0) \nonumber \\
 &\times& 
 V^{ef}[(\vec 0,T),(\vec 0,0)]
 \rho_0^{fg}(\vec 0, \vec z)
 G_q^{gh}[(\vec z,0),(\vec u,0)] \rho_0^{ha}(\vec u, \vec R) 
  \label{eq:BB_BB} \\
 &-&  V^{ab}[(\vec R,0),(\vec R,T)]
 \rho_T^{bc}(\vec R, \vec x) G_q^{cd}((\vec x,T),(\vec y,0))
 \rho_0^{da}(\vec y, \vec R) \nonumber \\
 &\times&
 V^{ef}[(\vec 0,T),(\vec 0,0)]
 \rho_0^{fg}(\vec 0, \vec z) G_q^{gh}((\vec z,0),(\vec u,T))
 \rho_T^{he}(\vec u, \vec 0) ~. \nonumber
\end{eqnarray}
We compute the two terms in eq.~(\ref{eq:BB_BB}) again with the
Gaussian random source method using two independent Gaussian sources,
$\eta$ and $\zeta$, one for each of the two light quark propagators:
\begin{eqnarray}
&& \langle O^\dagger_M(\vec 0, \vec R, T) O_M(\vec 0, \vec R, 0) 
 \rangle \nonumber \\
\label{eq:BB_BBr}
 &=& \langle \Psi_\eta^f(\vec 0,0) \eta^{\dagger a}(\vec R, 0)
 V^{ab}[(\vec R,0),(\vec R,T)] \rangle_\eta \langle
 \Psi_\zeta^b(\vec R,T) \zeta^{\dagger e}(\vec 0,T)
 V^{ef}[(\vec 0,T),(\vec 0,0)]
 \rangle_\zeta \nonumber \\
 &-& \langle \eta^{\dagger a}(\vec R, 0)
 V^{ab}[(\vec R,0),(\vec R,T)]
 \Psi_\eta^b(\vec R,T) \rangle_\eta \langle \zeta^{\dagger e}(\vec 0,T)
 V^{ef}[(\vec 0,T),(\vec 0,0)]
 \Psi_\zeta^f(\vec 0,0) \rangle_\zeta \nonumber ~.
\end{eqnarray}
Again, this quantity is averaged over the same choices of origin and
displacement as the Wilson loop.  It is important to note from our
expressions for the correlators that all can be computed using $O(N)$
methods for $N$ random sources.

%
\section{Transfer Matrix Applied to String Breaking}
\label{sec:tm_appendix}
%
We review the Sharatchandra-Thun-Weisz (STW) Fock-space formulation of
the staggered fermion partition function with specific application to
the operators used in our string breaking study \cite{ref:STW}.

\subsection{Introduction}
The fermion action is given by 
\begin{equation}
S  = \sum_{t} S_3(\bar q_t, q_t,U_t) + 
      \sum_{r,t} \alpha_{r,t;4}\left[\bar q_{r,t+1}U_{r,t}  q_{r,t} 
 - \bar q_{r,t}U^\dagger_{r,\hat t}  q_{r,t+1}\right] ~,
\end{equation}
where the spatial part of the action is
\begin{equation}
   S_3(\bar q_t, q_t,U_t) = \sum_r \left\{ 2m\bar q_{r,t}  q_{r,t}
    + \sum_i \alpha_{r,t;i}\left[\bar q_{r+\hat \imath,t} U_{r,t;i}  q_{r,t} 
   -  \bar q_{r,t} U^\dagger_{r,t;i}  q_{r+\hat \imath,t}\right]\right\} ~.
\end{equation}

With the STW phase convention
\begin{eqnarray}
   \alpha_{r,t;1}  &=& (-)^z \ \ \ \ \ \alpha_{r,t;2} = (-)^x \\
   \alpha_{r,t;3}  &=& (-)^y \ \ \ \ \ \alpha_{r,t;4} = (-)^{x+y+z} ~. 
  \nonumber
\end{eqnarray}
With this choice the phases have no $t$ dependence.  The imaginary
time variable ranges over $0,\ldots{},N-1$, and the antiperiodic
boundary condition requires $ q_{r,N} = - q_{r,0}$.

The partition function is given by
\begin{equation}
  Z = \int d\bar q d q \exp(S) ~,
\end{equation}
where $\bar  q$ and $ q$ denote the full set $\bar q_{r,t}$ and
$ q_{r,t}$.  We would like to convert the Grassmann integral into a
Fock-space operator trace.  To this end STW first eliminate the KS
phase in the time direction altogether, by changing variables to
$ q^\dagger = \alpha_{r,t;4}\bar q$.  The spatial action becomes
\begin{equation}
   S_3( q^\dagger_t, q_t,U_t) = 
    \sum_r \left\{ 2m q^\dagger_{r,t}  q_{r,t} \alpha_{r,t;4}
   + \sum_i \alpha^\prime_{r,t;i}\left[ q^\dagger_{r+\hat \imath,t} 
            U_{r,t;i}  q_{r,t} 
   -   q^\dagger_{r,t} U^\dagger_{r,t;i} 
       q_{r+\hat \imath,t}\right]\right\} ~,
\end{equation}
where
\begin{equation}
    \alpha^\prime_{r,t;i} = \alpha_{r,t;4}\alpha_{r,t;i} ~.
\end{equation}
Then STW introduce a dummy set of Grassmann variables $ p_t =
 q^\dagger_t$ and $ p^\dagger_t =  q_t$.  The action is then
\begin{equation}
S  = \sum_{t} S_3( q^\dagger_t, p^\dagger_t)/2 
    + S_3( p_t, q_t)/2 
    -  \left[  p^\dagger_t  p_{t+1} +  q^\dagger_t  q_{t+1} \right] ~.
\end{equation}
where we have suppressed the sum over spatial coordinate $r$.
The partition function now includes integration
over the dummy variables with a delta-function constraint:
\begin{equation}
  Z =  \int d q^\dagger d q d p^\dagger d p 
   \delta( p -  q^\dagger) \delta( q -  p^\dagger)\exp(S) ~.
\end{equation}
Here the delta function implies a product over delta functions on each
lattice site.  The Grassmann delta function is simply
\begin{equation}
   \delta( q -  p^\dagger) = ( q -  p^\dagger) ~.
\end{equation}
We introduce a Fock space by associating creation and annihilation
operators with each Grassmann variable.  The corresponding Fock space
operators are denoted by a hat.  It is convenient to introduce
Grassmann coherent states \cite{ref:coherent}:
\begin{eqnarray}
  \hat q\left| q\right\rangle &=&  q\left| q\right\rangle \mbox{\ \ \ \ \ }
   \hat p\left| p\right\rangle =  p\left| p\right\rangle  \nonumber \\
  \left\langle q^\dagger\right|\hat q^\dagger &=& 
     \left\langle q^\dagger\right| q^\dagger
  \mbox{\ \ \ \ \ }
  \left\langle p^\dagger\right|\hat p^\dagger = 
   \left\langle p^\dagger\right| p^\dagger ~.
\end{eqnarray}
The coherent states satisfy completeness and trace relations:
\begin{eqnarray}
  1 &=& \int d q^\dagger d q \left| q\right\rangle\left\langle q\right| 
     \exp(- q^\dagger q) \nonumber\\
  \mbox{Tr}{A} &=& \int d q^\dagger d q 
   \left\langle q^\dagger\right| A \left| q\right\rangle  
    \exp( q^\dagger q) ~.
\end{eqnarray}
With these identities we can reorganize the factors in $\exp(S)$
in the form
\begin{eqnarray}
  && \exp(S) 
   \delta( p -  q^\dagger) \delta( q -  p^\dagger)\nonumber \\
     && = \prod_t \left\{
    e^{- p^\dagger_t p_{t+1} -  q^\dagger_t q_{t+1}}
   \left\langle q^\dagger_t  p^\dagger_t\right|
    e^{S_3(\hat q^\dagger, \hat p^\dagger)/2}
    \,\,
     :(\hat p - \hat q^\dagger)(\hat q - \hat p^\dagger):
     \,\,
    e^{S_3(\hat q, \hat p)/2}
    \left| q_t  p_t\right\rangle \right\} ~.
 \nonumber
\end{eqnarray}
where $::$ denotes operator normal ordering.  Using the completeness
and trace identities and the antiperiodic boundary condition, we can
then write the partition function in terms of a transfer matrix operator:
\begin{equation}
  Z = \mbox{Tr} {\cal T}^N ~,
\end{equation}
where
\begin{equation}
   {\cal T} = e^{S_3(\hat q^\dagger,\hat p^\dagger)/2}
       \,\,
     :(\hat q^\dagger - \hat p)(\hat q - \hat p^\dagger):
    \,\,
       e^{S_3(\hat q,\hat p)/2} ~.
\end{equation}
This is a manifestly Hermitian, but not positive definite, transfer
matrix.
\subsection{Quantum mechanics}
To see how this transfer matrix works, as a warmup exercise, suppose
there is only one spatial site (quantum mechanics).  The transfer matrix
is then
\begin{equation}
  {\cal T} = (1 + m \hat q^\dagger\hat p^\dagger)  
      :(\hat q^\dagger - \hat p)(\hat q - \hat p^\dagger):
      (1 - m \hat q \hat p) ~.
\end{equation}
It has eigenstates ($m = \sinh \gamma$)
\begin{equation}
  \hat q^\dagger \left|0\right\rangle, \,\,\, 
  \hat p^\dagger \left|0\right\rangle, \,\,\,
  \left|\pm c\right\rangle = (1 \mp e^{\mp \gamma} 
    \hat q^\dagger\hat p^\dagger) 
   \left|0\right\rangle/ \sqrt{1 + e^{\mp 2 \gamma}}
\end{equation}
with eigenvalues $1, -1, \pm e^{\mp \gamma}$.  Thus the ground state
is $\left|-c\right\rangle$.  The ``degenerate'' $ p$ and $ q$ states
are interpreted as single particle and single antiparticle states, and
the state $\left|+c\right\rangle$, as a state with a particle and
antiparticle pair.  The partition function is ($N$ even)
\begin{equation}
  Z = 2 + 2 \cosh(N\gamma) ~.
\end{equation}

Next we consider the propagator for a single massive fermion, created
at time slice 0 and propagating to time slice $T$.
\begin{equation}
  G(T) = \int d q^\dagger d q \exp(S) 
    q_T q^\dagger_0/ Z ~.
\end{equation}
Converting to the Fock space basis as we did with the partition
function leads to
\begin{equation}
  G(T) = \mbox{Tr} [{\cal T}^{N-T-1} 
     \hat p^\dagger {\cal T}^{T+1} \hat p] /Z  ~.
\end{equation}
This example shows that a $T+1$ power is natural.
If we assume that $N$ is large, so only the ground state contributes
to the traces, we have ($T \ll N$)
\begin{equation}
  G(T) = e^{-\gamma T}/[2\cosh(\gamma)]
\end{equation}
with $\gamma$ interpreted as the energy of the propagating state.  For
large mass the correlator is approximately $1/(2m)^{(T+1)}$.

The antiparticle propagator (with $ q^\dagger_T q_0$ instead) is
also easily computed with the result
\begin{equation}
     G(T) = -(-)^T e^{-\gamma T}/[2\cosh(\gamma)] ~.
\end{equation}

Next, we introduce a second heavy flavor of mass $M$ ($M = \sinh
\Gamma$), denoted by $ h$.  We consider the propagation of a
heavy-light meson.  For simplicity we use an interpolating operator
with a local (point) wave function. In this case the Grassmann
integration involves the combination $ q^\dagger_T h_T
 h^\dagger_0 q_0$.  Following the same steps as before leads to
\begin{equation}
  G_B(T) = -(-)^T e^{-(\gamma+\Gamma)T}/[4\cosh(\gamma)\cosh(\Gamma)] ~.
\end{equation}

\subsection{String-breaking operators}

Here we construct the operators required for our string-breaking
study.  Two interpolating operators, $O^\dagger_F$ and $O^\dagger_M$,
are used, one that creates a static quark-antiquark pair with a
connecting flux tube and another that creates a pair of static-light
mesons.  A simple way to construct the flux tube operator is to
introduce a new heavy flavor $ H$ of mass $M_H$.  Our
conventions for the on-axis (direction $\hat \imath$) string-creation
operator can be obtained from the Grassmann product $(2M_H)^R
\bar h_{0,t}  H_{0,t} \bar H_{R,t}  h_{R,t}$ after integrating
out the new flavor in the static limit $M_H \rightarrow \infty$.
The result is a heavy quark-antiquark creation operator connected by
the static quark propagator.  In the Grassmann basis it is
\begin{equation}
  O^\dagger_F(R,t) = F^\dagger(R,t) = \bar  h_{0,t}
     \left[\prod_{r=0}^{R-1} (-\alpha_{r,t;i} U^\dagger_{r,t;i})\right]
      h_{R\hat \imath,t} ~.
\end{equation}
Our corresponding annihilation operator is similarly generated from
the product $(2M_H)^R \bar  h_{R,t}  H_{R,t} \bar H_{0,t}
 h_{0,t}$, yielding
\begin{equation}
  O_F(R,t) = F(R,t) = \bar h_{R \hat \imath,t}
     \left[\prod_{r=0}^{R-1} \alpha_{r,t;i} U_{r,t;i}\right]
      h_{0,t} ~.
\end{equation}
(With all of our static quark propagators, it is convenient to
introduce a separate, heavy flavor for each straight-line segment.)
The meson-antimeson creation and annihilation operators are simply
\begin{eqnarray}
  O^\dagger_M(R,t) &=& \bar B(0,0) B(R,0) = \bar  h_{0,t}  q_{0,t} 
                      \bar q_{R\hat \imath,t}  h_{R \hat \imath,t} 
            \nonumber \\
  O_M(R,t)         &=& \bar B(R,t) B(0,t) = \bar  h_{R\hat \imath,t} 
                        q_{R \hat \imath,t} 
                       \bar q_{0,t}  h_{0,t} ~.
\end{eqnarray}
We are interested in the two-channel correlators in the static limit
$M_h \rightarrow \infty$ with the heavy quarks fixed at $0$ and $R$:
\begin{equation}
  G_{AB}(R,T) = \left\langle O_A(R,T)O^\dagger_B(R,0) \right\rangle
       (2M_h)^{2T} ~,
\end{equation}
where $A,B \in \{M,F\}$.  The conversion to Fock space proceeds as
before.  With $\delta(g - h^\dagger) \delta(h - g^\dagger)$ the Fock
space operators are
\begin{eqnarray}
  \hat O^\dagger_F(R) &=&  \hat g_{0,t}\alpha_{0,t;4}
       \left[\prod_{r=0}^{R-1} (-\alpha_{r,t;i} U^\dagger_{r,t;i})\right]
     \hat h_{R\hat \imath,t}. \nonumber\\
  \hat O_F(R) &=& \hat h^\dagger_{R \hat \imath}\alpha_{R \hat \imath,t;4}
                  \left[\prod_{r=0}^{R-1}(\alpha_{r,t;i} U_{r,t;i})\right]
                  \hat g^\dagger_{0,t} \\
  \hat O^\dagger_M(R) &=& \hat g_{0,t} \alpha_{0,t;4} \hat q_{0,t} 
                         \hat p_{R\hat \imath,t}\alpha_{R\hat \imath,t;4} 
                         \hat h_{R \hat \imath,t} \nonumber\\
  \hat O_M(R) &=& \hat h^\dagger_{R\hat \imath,t}\alpha_{R\hat \imath,t;4}
                          \hat p^\dagger_{R \hat \imath,t} 
                          \hat q^\dagger_{0,t} \alpha_{0,t;4}
                          \hat g^\dagger_{0,t} ~. \nonumber
\end{eqnarray}
We recall that $\alpha_{r,t;4}$ is independent of $t$, and $\alpha_{R\hat
i,t;4}\alpha_{0,t;4} = (-)^R$, so the dagger indeed denotes the Fock
space hermitian conjugate.  The desired correlators are then
\begin{equation}
 G_{AB}(R,T) = \left\langle \rm{vac}\right|O_A(R)
        {\cal T}^{T+1}O^\dagger_B(R) \left| \rm{vac} \right\rangle ~.
\end{equation}
An eigenstate $\left|n\right\rangle$ of the transfer matrix with eigenvalue
$\lambda_n$ contributes 
\begin{equation}
   G_{ABn}(R,T) = Z_{An}^*(R) Z_{Bn}(R)[\lambda_n(R)]^{T+1} ~,
\end{equation}
where
\begin{equation}
  Z_{An}(R) = \left\langle n,R\right| \hat O_A(R) \left|vac\right\rangle ~.
\end{equation}
This result is the basis for Eq.~(\ref{eq:transition}).

\subsection{Staggered spin and flavor considerations}
The external states in our analysis are built from two operators: one
that creates a static quark-antiquark pair at separation $R$ and one
that creates a pair of static quark-light quark mesons at separation
$R$.  Both the static quark and light quark carry four continuum
flavors.  One may ask, in the continuum limit, what spin, parity, and
flavor combinations occur? 

\subsubsection{Flavors and spins of the static-light meson}

Golterman and Smit \cite{ref:GS} give an analysis of the flavor
content of staggered quark-antiquark mesons.  Our static-light meson
wavefunction has a local component (zero displacement) and a component
with displacement 2 along any axis.  An even displacement does not
change the light-quark flavor and spin wavefunction, so for simplicity
we analyze only the local component,
\begin{equation}
  \bar h(2\vec y + \vec \eta,t) q(2\vec y + \vec \eta,t) ~,
\end{equation}
where the c.m.\ offset $\vec \eta$ ranges over the eight sites in a
unit cube.  

The more familiar flavor-spin content is displayed in the notation of
Kluberg-Stern {\it et al.}  \cite{ref:Kluberg-Stern:1983dg}:
\begin{equation}
  q(2\vec y + \vec \eta,t) = \frac{1}{2}
        \Gamma^{\alpha a}_{\eta} q^{\alpha a}(\vec y)
\end{equation}
where a sum over four flavor $\alpha$ and four spin $a$ indices is
assumed, and 
\begin{equation}
  \Gamma^{\alpha a}_{\eta} = \gamma_0^{\eta_0}\gamma_1^{\eta_1}
      \gamma_2^{\eta_2}\gamma_3^{\eta_3}.
\end{equation}
A similar expression holds for the heavy quark operator $h(2\vec y +
\eta,t)$.  We have ignored the gauge connection in these expressions.
The coefficients $\Gamma^{\alpha a}_{\eta}$ give the flavor-spin
wavefunction of a quark created at site $\eta$.

The zero-three-momentum projection of the local static-light operator
is given by
\begin{equation}
  B_{\vec \eta}(t) = \sum_{\vec y} \bar h(2\vec y + \vec \eta,t) 
        q(2\vec y + \vec \eta,t) ~.
\end{equation}
All eight operators, corresponding to the eight values of $\vec \eta$,
belong to Golterman's class 0.  Because the static quark must
propagate in place, the correlation matrix for these eight states is
diagonal in and independent of $\vec \eta$.  Apart from a volume
factor, it is exactly the static-light correlator we have computed.
Linear combinations of these operators with coefficients differing
only by signs yield operators belonging to the rest-frame symmetry
group of the discrete transfer matrix.  For example, the operator
belonging to the one-dimensional representation ${\bf 1}^{+-}$ in the
Golterman-Smit notation is given by
\begin{equation}
  B_{{\bf 1}^{+-}}(t) = \frac{1}{\sqrt{8}}
    \sum_{\vec \eta}(-)^{a_1+a_2+a_3} B_{\vec \eta}(t) ~.
\end{equation}
In the Kluberg-Stern {\it et al.} basis this operator is written as
\begin{equation}
  \frac{1}{8}\sum_{\vec y} \bar h(y) [\gamma_5 \times \gamma_5 \pm 
            \gamma_0 \times \gamma_0] q(y)
\end{equation}
where the first gamma factor in the tensor product operates on the
spin basis and the second, the flavor basis, and the sign is plus for
even $t$ and minus for odd.

This operator generates both a pion-like and sigma-like state in the
continuum, in which, respectively, the light quark is found in an $S$
or $P$ orbital around the static quark.  Similar linear combinations
yield eight operators, altogether belonging to the irreps ${\bf
1^{++}}$, ${\bf 1^{+-}}$, ${\bf 3}^{\prime\prime\prime\prime++}$, and
${\bf 3}^{\prime\prime\prime\prime++}$, each interpolating one of
eight $S$-wave channels: two pion-like ($B$-mesons) and two rho-like
($B^*$-mesons).  Each of these is paired with one of eight $P$-wave
channels: two sigma-like and two pseudovector-like, respectively.  All
$S$-wave states are degenerate, as are all $P$-wave states.  A fixed
offset $\vec \eta$ corresponds to a distinct linear combination of
these degenerate states.  Alternatively, we may say that a fixed
offset $\vec \eta$ corresponds to the pairing of a single flavor-spin
species of light quark and heavy quark.

Thus our two-meson operator interpolates all combinations of $B$ and
$B^*$ and their $P$-wave counterparts.  Our procedure for computing
correlators sums over all cubic rotations of $\vec R$ at fixed $|R|$,
so is consistent with zero total angular momentum for the two-meson
and flux-tube states.  Transitions of the type $SS \rightarrow SP$ are
accompanied by a change in orbital angular momentum.

Static-light mesons in the Golterman classes 1, 2, and 3 are similarly
degenerate within each class with multiplicities 24, 24, and 8,
respectively.  One would expect in analogy with the light $\bar q q$
mesons that on a coarse lattice the energies of the class nonzero
static-light mesons are higher than that of class 0.  In the continuum
limit all 64 become degenerate because of flavor and heavy quark
symmetry.  A multiplicity factor of 16 comes from separate
heavy-flavor and light-flavor symmetry and a multiplicty factor of $4
= 1+3$ for degenerate singlet and triplet spin combinations.

\subsubsection{Flavors and spins of the static-static meson}

The flux-tube operator is designed so that it belongs to the same
representation of the symmetry group of the transfer matrix as the
meson-antimeson operator, thus permitting mixing of the states they
create.  The zero momentum projection of the meson-antimeson operator
at any separation $R$ is in Golterman class 0.  At $R = 0$ the
flux-tube operator creating a static quark-antiquark pair is trivially
one of eight Golterman class 0 operators.  For it to remain in this
class as the quarks separate, one must include the hopping phases
$\pm\alpha_\mu(x)$ with the gauge-link operators that excite the
electric flux accompanying the creation of the static pair
(\ref{eq:link_with_phases}).  A simple way to see this is to consider
that the operator class is a symmetry of the transfer matrix, so
acting upon the zero separation state with the heavy-quark hopping
matrix, which incorporates the phases, preserves the operator class.
The eight operators consist of two pion-like and six rho-like
operators.  More specifically, the static-quark-antiquark operator at
all $R$ is a linear combination of the two local (heavy quark)
pseudoscalars $\pi_5$ and $\pi_{05}$ and the six local vector mesons
$\rho_i$ and $\rho_{i0}$.  In the $b$-quark system, these states would
be classified as the $\Upsilon$ and the $\eta_b$.

In free (static) propagation the even parity $P$-wave partners of both
are trivially absent in the static limit.  On the other hand, in the
presence of dynamical staggered quarks and in the strong-coupling
limit, we see no reason that dynamical quark loops could not generate
an anomalous, opposite-parity partner in the Wilson loop correlator at
second order in the mixing coefficients.  Thus, the appearance of both
oscillating and nonoscillating components in the Wilson loop would be
a signal of string-breaking at strong coupling.  In the continuum
limit the anomalous components should vanish, in keeping with the
expectation that lattice staggered and nonstaggered fermion actions
have identical limits.  On our moderately coarse lattices, in the
transition correlator connecting the flux-tube and meson-antimeson
channels, the signal appears in leading order, but from our mixing
analysis, we find that the anomalous parity component is smaller.

\subsubsection{Light flavor artifacts in the meson-meson correlator}

The peculiarities of staggered flavor symmetry require paying special
attention to light quark flavor counting.  Two issues confront us.
First, in our meson-antimeson channel, in the continuum limit the
light quarks can combine as a flavor singlet or flavor non-singlet
with, in principle, distinct energies.  Since the flux-tube state is
necessarily a light-flavor singlet, only the singlet components mix,
leaving the nonsinglet as a potential additional spectral component.
Second, the gauge configurations were generated in the presence of
$N_f = 2$ flavors, through the usual device of taking the square root
of the intrinsically four-flavor fermion determinant.  However, the
light quarks in our source and sink carry four flavors.  A mismatch
between the number of valence and sea quarks gives rise to additional
spectral components in the disconnected meson-to-meson correlator.

To count flavors, we construct a toy model similar to the mixing model
of subsection \ref{sec:mixingmodel}.  However, for present purposes we
ignore negative transfer matrix eigenvalues and deal directly with the
Laplace transform of the correlators.  Further, it is sufficient to
concentrate on the meson-meson correlator $G_{MM}$.

We begin with the disconnected meson-meson correlator $G_{MMD}$ at a
fixed separation $R$.  As we have noted above, the static-light mesons
are created in Golterman class 0.  In the absence of glueball exchange
between the meson and antimeson, they remain in class 0.  Glueball
exchange can excite class $n>0$ levels in pairs.  Such an effect is
presumably short range, because the glueball mass is of order one GeV.
Thus, on a coarse lattice we expect four distinct energy levels, and
the Laplace transform of the $G_{MMD}$ correlator has the form
\begin{equation}
  G_{MMD} = \sum_{n=0}^3\frac{Z_n^2}{E - E_n}.
\end{equation}
At least for $R$ larger than a few tenths of a fermi we expect the
lowest level to have the largest overlap with the source and sink so
that $Z_0 >> Z_n$ for $n>0$.  So we simplify our model, writing
\begin{equation}
  G_{MMD} = \frac{Z^2}{E - E_0} + \ldots{}.
\end{equation}
where the remaining terms are higher energy and weakly coupled,
except, possibly, at small $R$.

In the continuum limit flavor symmetry requires only two energy
levels, depending on whether the light quarks form a flavor singlet or
nonsinglet.  Thus we use $E_0$ to denote the singlet and $E_1$, the
nonsinglet.  Since our source/sink operator generates a single
flavor-spin species of light quark and light antiquark, for four light
flavors the relative weight of singlet to nonsinglet is $1:3$, and we
get
\begin{equation}
  G_{MMD}|_{\rm cont} = \frac{Z^2}{E - E_0} + 3 \frac{Z^2}{E - E_1}.
\end{equation}

Next, we consider the connected correlator $G_{MMC}$.  We treat light
quark pair annihilation and creation as a weak process with amplitude
$x$ and analyze this correlator as a perturbation series in $x$.  Thus
the leading term in the connected diagram has the form
\begin{equation}
  G_{MMC}^{(2)} = \frac{Z^2 x^2}{(E-E_0)^2(E-E_f)}
\end{equation}
where $E_f$ is the energy of the flux-tube state.  The amplitude $x$
gives the weight for annihilation or creation of a single light
quark/antiquark flavor.  Here is the sequence of events: The source
state couples to the lowest level, which propagates with energy $E_0$.
It then annihilates with amplitude $x$ to produce the flux-tube state,
propagating with energy $E_f$.  Pair creation leads back to the lowest
level, which propagates with energy $E_0$ before coupling to the sink.

The next higher order contribution to $G_{MMC}$ comes from a
quark-loop insertion.  The sequence of events is the same as in the
leading-order contribution, except that an extra pair creation and
annihilation process occurs in the flux-tube state.  Pair creation on
a coarse lattice leads to any of the four levels $E_i$ at the same
order in $x$.  In the continuum limit and large $R$ these levels are
degenerate, and would each count with weight $x^2 N_f/4$, the factor
$N_f/4$ arising from the fermion determinant.  For simplicity, to
model the full effect of the internal quark loop, we assign it a
weight $x^{\prime 2}$ with $x^{\prime 2} = x^2 N_f$ in the continuum
and represent only the lowest level $E_0$.  Thus we have
\begin{equation}
  G_{MMC}^{(4)} = G^{(2)}_{MMC}
   \frac{x^{\prime 2}}{(E - E_0)(E - E_f)}.
\end{equation}
Continuing with these simplifications to all orders, we may sum the
perturbation series to obtain $G_{MM} = G_{MMD} + G_{MMC}$, where
\begin{equation}
   G_{MM} = Z^2 \frac{1 + (x^2 - x^{\prime 2})/[(E - E_0)(E - E_f)]}
    {E - E_0 - x^{\prime 2}/(E - E_f)} + \ldots{}.
\end{equation}
The ellipsis represents contributions to the disconnected diagram from
light-flavor nonsinglet terms.

This toy flavor-counting model shows the desired shifted pole at $E_0
+ x^{\prime 2}/(E - E_f)$, but additional, complicating spectral components at
$E_0$, $E_f$, and the light-flavor nonsinglet level(s).  These
additional components are an artifact of our choice of source and
sink.  While, in principle, one may carry through the spectral
analysis of the correlators, remembering to include the artifacts, the
analysis would be unnecessarily complicated.  For the present
numerical study, weak mixing and strong flavor symmetry breaking on
our coarse lattice render the extra spectral components harmless.
First, strong coupling breaks flavor symmetry and lifts the levels
$E_1,\ldots{}$ above $E_0$, making it less likely to confuse them with
$E_0$.  Second, because of weak mixing, we are unable to detect the
internal quark loop directly.  Thus we measure the mixing parameter
$x$, but do not see $x^\prime$.

Closer to the continuum, it is desirable to replace the meson-meson
interpolating operator with a better one, building in an explicit
projection of the light-quark flavor singlet.  An example is a state
formed from the tensor product of a $\sigma$ meson and a static
quark-antiquark flux-tube.  Clearly, such a projection eliminates the
nonsinglet contribution.  But to eliminate the other artifacts also
requires $x^2 = x^{\prime 2}$ in the toy model, which is not automatic
when there are two sea quark flavors but four source and sink
flavors.  The remaining artifacts are eliminated by adjusting the
weight of the disconnected diagram, relative to the connected diagram.
To be precise, once a projection to an SU(4) flavor singlet has been
done in the state $M$, the proper weighting is $G_{MMD} + N_f/4
G_{MMC}$.

\begin{table}
\caption{Fit ranges in $t$, chi square, potentials, and $\kappa$ vs $R$.
Jackknife errors are given.
\label{tab:resultsa}
}
\begin{tabular}{lcccddddl}
    R   & $G_{FF}$ & $G_{FM}$ & $G_{MM}$ & $\chi^2/df$ & $V_1(R)$ 
   & $V_2(R)$ & $V_3(R)$ \\
\hline
1.0 & [4,12] & [1,12] & [2,12] & 20.0/23 &  1.57(3) &  1.71(10) &  0.5229(4) \\
2.0 & [4,11] & [1,10] & [2,10] & 24.0/18 & 1.62(5) & 1.91(13) &  0.8181(15) \\
2.83 & [3,10] & [1,9] & [2,9] & 11.1/16  & 1.61(3) & 1.81(8) &  1.0066(13) \\ 
3.0 & [4,9] & [1,9] & [2,9] & 23.4/14    & 1.63(5) & 1.91(13) &  1.024(4) \\  
4.0 & [4,9] & [1,9] & [2,9] & 16.2/14    & 1.61(6) & 1.85(13) &  1.198(10) \\ 
5.0 & [2,9] & [1,8] & [2,8] & 18.6/14    & 1.63(5) & 1.88(11) &  1.353(3) \\  
5.66 & [2,9] & [1,7] & [2,7] & 14.6/12   & 1.59(4) & 1.74(9) &  1.458(2) \\   
6.0 & [2,9] & [1,7] & [2,7] & 14.8/12    & 1.54(4) & 1.69(10) &  1.507(4) \\  
7.0 & [2,8] & [2,7] & [2,7] & 9.6/10     & 1.69(4) & 1.96(12) &  1.648(14) \\ 
8.0 & [2,8] & [2,6] & [2,7] & 13.3/9     & 1.61(4) & 1.81(6) &  1.78(6) \\    
9.0 & [2,8] & [2,5] & [2,6] & 9.6/7      & 1.49(3) & 1.54(8) &  1.936(14) \\  
10.0 & [2,8] & [2,5] & [2,6] & 8.7/7     & 1.58(5) & 1.79(14) &  2.09(2) \\   
\end{tabular}
\end{table}
\begin{table}
\caption{Couplings vs $R$. Odd and even $R$ values are grouped. (The
displacements with $R = \sqrt{8}$ and $2\sqrt{8}$ have even Cartesian
components.)
\label{tab:resultsb} }
\begin{tabular}{ldldldl}
$R$ & $Z_{F1}$ & $Z_{M1}$& $Z_{F2}$ & $Z_{M2}$& $Z_{F3}$ & $Z_{M3}$ \\
\hline
1.0 &  0.162(3) &  14.0(1.0) & -0.002(3)  &  13(2) & -1.2733(10) &  3.356(3) \\
3.0 &  0.24(3)  &  16(2)     &  0.005(4)  &  19(5) & -1.582(14)  &  4.33(7) \\
5.0 &  0.23(3)  &  16(2)     &  0.005(3)  &  18(4) & -1.784(6)   &  2.9(2) \\
7.0 &  0.9(8)   &  16(5)     & -0.00(2)   &  21(5) & -1.8(4)     &  8(6) \\
9.0 & -0.01(3) &   12.1(7)   & -0.007(10) &  9.2(1.4) & -2.13(5) &  0.2(7) \\
\hline
2.0  &  0.0180(14) &  16(2)     &  0.005(7)  & 19(5) & -1.443(5) &  1.970(8) \\
2.83 &  0.0160(7)  &  15.4(1.1) &  0.051(4)  & 16(2) & -1.540(3) &  1.577(9) \\
4.0  &  0.0141(12) &  16(2)     &  0.056(15) & 17(4) & -1.70(4)  &  1.19(3) \\
5.66 &  0.0087(11) &  14.9(1.3) &  0.06(2)   & 14(2) & -1.780(5) &  0.71(7) \\
6.0  &  0.0067(13) &  13.4(1.2) &  0.12(6)   & 12(2) & -1.890(9) &  1.0(3) \\
8.0  & -0.003(12)  &  15.5(1.3) & -1(2)      & 10(20) &  1(2)    &  9(15) \\
10.0 & -0.014(10)  &  15(2)     &  0.2(2)    & 15(4) & -2.24(6)  &  3(3) \\
\end{tabular}
\end{table}
\begin{table}
\caption{Results of mixing model analysis based on a three-exponential fit 
showing the coefficient ratio $d^\prime$ prior to imposing the 
{\it a posteriori} sum rule constraint and $d$ after, the increase in $\chi^2$, 
and the mixing parameters $x$  and $y$ vs $R$.  This mixing model seems 
plausible only for $R \ge 5.66$.  Jackknife errors are given.  Values for
odd and even $R$ are grouped as in Table \protect\ref{tab:resultsb}.
\label{tab:mixing_three} 
}
\begin{tabular}{lddddd}
    R   & $d^\prime$ & $\Delta\chi^2$ & $d$ & $x$       & $y$  \\
\hline
 1.00 &   0.52(3) & 520.    &   0.68(6) &      0.056(3) &     -0.019(3) \\
 3.00 &   0.56(3) & 200.    &   0.50(9) &      0.023(3) &     -0.007(2) \\
 5.00 &   0.72(2) & 98.     &   -0.0(8) &      0.000(6) &    -0.0010(9) \\
 7.00 &  1.00(12) & 0.0     & 1.0001(13) &     0.004(2) &    -0.0004(8) \\
 9.00 &     -1(9) & 0.21    & 1.002(10) &   0.00090(15) &    -0.0005(5) \\
\hline
 2.00 &   0.14(3) & 1300.   &   2.4(12) &      0.002(3) &     -0.036(8) \\
 2.83 &   0.43(5) & 230.    &  0.87(10) &    -0.0044(4) &   -0.0077(11) \\
 4.00 &  0.58(10) & 23.     &  1.08(10) &    -0.0030(6) &   -0.0075(15) \\
 5.66 &   0.79(9) & 12.     &   1.00(3) &    -0.0016(2) &    -0.0029(3) \\
 6.00 &   0.85(7) & 5.8     &    0.9(2) &    -0.0009(3) &   -0.0029(10) \\
 8.00 &    1.2(4) & 1.6     &    1.0(6) &      -0.00(2) &      0.01(15) \\
10.00 &    0.3(2) & 2.1     &    0.9(3) &     0.0003(4) &    -0.0006(5)
\end{tabular} 
\end{table}
\begin{table}
\caption{Same as table \protect\ref{tab:mixing_three} but with a
four-exponential fit, constraining $V_1(R) = 2E_S$ and  $V_4(R) = 2E_P$.
The mixing model now seems plausible for $R \ge 2$.
Jackknife errors are given.
\label{tab:mixing_four} }
\begin{tabular}{lddddd}
    R   & $d^\prime$ & $\Delta\chi^2$ & $d$ & $x$       & $y$  \\
\hline
 1.00 &  0.52(5) & 140.    &    1.0(9) &      0.10(14) &    0.007(4)  \\ 
 3.00 &   0.4(2) & 19.     &   0.94(2) &      0.051(2) &    0.0089(13) \\ 
 5.00 &   0.6(4) & 1.8     &   0.99(2) &     0.0139(6) &     0.0034(7) \\
 7.00 &     1(2) & 0.00    &  1.001(9) &     0.0032(6) &     -0.001(4) \\ 
 9.00 &    0(20) & 0.032   &    0.9(6) &       0.2(15) &       -0.1(7) \\ 
\hline
 2.00 &   0.5(2) & 5.8     &   1.06(4) &     -0.037(3) &     -0.016(2) \\ 
 2.83 & 0.61(15) & 9.6     &  1.003(2) &   -0.0200(11) &    -0.0129(3) \\ 
 4.00 &   0.9(4) & 0.037   & 1.0000(2) &     -0.010(2) &    -0.0075(3) \\ 
 5.66 &   0.3(5) & 4.8     &   0.96(4) &   -0.0038(11) &    -0.0025(3) \\ 
 6.00 &   0.4(6) & 0.83    &  1.001(2) &   -0.0011(13) &    -0.0027(3) \\ 
 8.00 &   1.5(3) & 5.5     &  1.1(16)  &      0.007(6) &   -0.0008(14) \\ 
10.00 &   0.3(2) & 1.6     &      1(9) &      0.002(5) &    -0.001(13)
\end{tabular} 
\end{table}


\begin{references}
\bibitem{ref:quenchedpot} 
         G.~Bali and K.~Schilling, 
		Phys.\ Rev.\ D {\bf 46}, 2636 (1992);
		{\bf 47}, 661 (1993);
         S.P.~Booth {\it et al.} (UKQCD Coll.), 
		Phys.\ Lett.\ B {\bf 294}, 385 (1992);
         Y.~Iwasaki {\it et al.}, 
		Phys.\ Rev.\ D {\bf 56}, 151 (1997);
         B.~Beinlich {\it et al.}, Eur.\ Phys.\ J.\ C {\bf 6}, 133 (1999),
		[hep-lat/9707023];
         R.G.~Edwards, U.M.~Heller and T.R.~Klassen, 
		Nucl.\ Phys.\ {\bf B517}, 377 (1998).
\bibitem{ref:fullpot1} 
         K.D.~Born {\it et al.}, 
		Phys.\ Lett.\ B {\bf 329}, 325 (1994);
         U.M.~Heller {\it et al.}, 
		Phys.\ Lett.\ B {\bf 335}, 71 (1994);
         U.~Gl\"assner {\it et al.} (SESAM Coll.), 
		Phys.\ Lett.\ B {\bf 383}, 98 (1998);
         C.~Bernard {\it et al.} (MILC Coll.), 
		Phys.\ Rev.\ D {\bf 56}, 5584 (1997); 
         S.~Aoki {\it et al.} (CP-PACS Coll.), 
		Nucl.\ Phys.\ B (Proc.\ Suppl.) {\bf 63}, 221 (1998).
         S.~Tamhankar and S.~Gottlieb, Nucl.\ Phys.\ B (Proc.\ Suppl.)
                {\bf 83-84}, 212 (2000).
         M.~Talevi {\it et al.} (UKQCD Coll.), 
		Nucl.\ Phys.\ B (Proc.\ Suppl.) {\bf 63}, 227 (1998).
\bibitem{ref:fullpot2} 
         B.~Bolder {\it et al.},
                Phys.\ Rev.\ D {\bf 63}, 074504 (2001)
                [hep-lat/0005018].
%
\bibitem{ref:Z2noise} It has been argued that Z(2) noise methods are superior.
S.~Dong and K.~Liu, Phys.\ Lett.\ B {\bf 328}, 130 (1994)
%
\bibitem{ref:michael} C. Michael, 
Nucl.\ Phys.\ B (Proc.\ Suppl.) {\bf 26} (1992) 417.
%
\bibitem{ref:DKKL}C.~DeTar, O.~Kaczmarek, F.~Karsch and E.~Laermann,
Phys.\ Rev.\ {\bf D 59}, 031501 (1999).
%
\bibitem{ref:drummond}
I.~T.~Drummond,
Nucl.\ Phys.\ B (Proc.\ Suppl.) {\bf 73}, 596 (1999).
%
\bibitem{ref:drummond2} 
I.~T.~Drummond, Phys.\ Lett.\ B {\bf 434}, 92 (1998).
%
\bibitem{ref:Knechtli_Sommer}
F.~Knechtli and R.~Sommer  [ALPHA collaboration],
Phys.\ Lett.\ B {\bf 440}, 345 (1998) [hep-lat 9807022];
Nucl.\ Phys.\ B {\bf 590}, 309 (2000)
[hep-lat/0005021].
%
\bibitem{ref:Stephenson}
P.~W.~Stephenson, Nucl.\ Phys.\ {\bf B550}, 427 (1999).
%
\bibitem{ref:deForcrand_Philipsen}
P.~de Forcrand and O.~Philipsen,
Phys.\ Lett.\  {\bf B475}, 280 (2000).
[hep-lat/9912050].
%
\bibitem{ref:Philipsen_Wittig}
O.~Philipsen and H.~Wittig,
Phys.\ Rev.\ Lett.\  {\bf 81}, 4056 (1998). 
%
\bibitem{ref:Trottier}
H.~Trottier, Phys.\ Rev.\ {\bf D 60}, 034506 (1999)
%
\bibitem{ref:Stewart_Koniuk}
C.~Stewart and R.~Koniuk, 
Phys.\ Rev.\  {\bf D 59}, 114503 (1999). 
%
\bibitem{ref:lacock_lat99}
C.~DeTar, U.~Heller, and P.~Lacock, Nucl.\ Phys.\ B (Proc.\ Suppl.)
{\bf 83}, 310 (2000).
%
\bibitem{ref:Pennanen_Michael}
P.~Pennanen and C.~Michael  [UKQCD Collaboration],
``String breaking in zero-temperature lattice QCD,'' [hep-lat/0001015].
%
\bibitem{ref:DET}
A.~Duncan, E.~Eichten, and H.~Thacker, 
``String Breaking in Four Dimensional Lattice QCD'', 
[hep-lat/0011076].
%
\bibitem{ref:SOMMER}
R.~Sommer, Nucl.\ Phys.\ {\bf B411}, 839 (1994).
%
\bibitem{ref:APEblock}
M.~Falcioni, M.~Paciello, G.~Parisi, and B.~Taglienti,
Nucl.\ Phys.\ {\bf B251}, 624 (1985).
M. Albanese {\it et al.} Phys. Lett. B {\bf 192}, 163 (1987).
%
\bibitem{ref:STW}
H.S.~Sharatchandra, H.J.~Thun, and P.~Weisz, 
Nucl.\ Phys.\ {\bf B192}, 205 (1981).
%
\bibitem{ref:Aoki}
S.~Aoki,
Nucl.\ Phys.\ Proc.\ Suppl.\ {\bf 94}, 3 (2001)
%
\bibitem{ref:Drummond_Horgan}
I.~T.~Drummond and R.~R.~Horgan,
Phys.\ Lett.\ B {\bf 447}, 298 (1999)
%
\bibitem{ref:coherent}
M.~Creutz,
Phys.\ Rev.~\ {\bf D 15}, 1128 (1977); 
D.~Soper, Phys.\ Rev.~\ {\bf D 18}, 4590 (1978).
%
\bibitem{ref:Kluberg-Stern:1983dg}
H.~Kluberg-Stern, A.~Morel, O.~Napoly and B.~Petersson,
Nucl.\ Phys.\  {\bf B220}, 447 (1983).
%
\bibitem{ref:GS}
M.~Golterman and J.~Smit, Nucl.\ Phys.\ {\bf B245}, 61 (1984);
M.~Golterman, Nucl.\ Phys.\ {\bf B273}, 663 (1986).
\end{references}
\end{document}